\documentclass[traditabstract]{aa} 

\usepackage{graphicx}
\usepackage{txfonts}
\usepackage{natbib}
\usepackage{longtable,lscape}
\usepackage{ulem}

\begin{document}

\title{Abundance analysis of the outer halo globular cluster Palomar~14
\thanks{Based on observations collected at the European Southern
  Observatory, Chile (Program IDs 077.B-0769). Table
  A.1 and A.2 is only available in electronic form at the CDS via anonymous
  ftp to cdsarc.u-strasbg.fr (130.79.125.5) or via
  http://cdsweg.u-strasbg.fr/Abstract.html.}}

\author{
  \c{S}eyma \c{C}al{\i}\c{s}kan\inst{1,2}
  \and Norbert Christlieb\inst{1}
  \and Eva K. Grebel\inst{3}
}

\institute{
  Zentrum f\"{u}r Astronomie der Universit\"{a}t Heidelberg,
  Landessternwarte, K\"{o}nigstuhl 12, 69117 Heidelberg, Germany\\
  \email{[nchristlieb/scaliskan]@lsw.uni-heidelberg.de}           
  \and 
  Department of Astronomy and Space Sciences, Ankara University, 
  06100, Tando\u{g}an, Ankara, Turkey\\
  \email{scaliskan@lsw.uni-heidelberg.de}
  \and
  Zentrum f\"{u}r Astronomie der Universit\"{a}t
  Heidelberg, Astronomisches Rechen-Institut,
  M\"{o}nchhofstr. 12--14, 69120 Heidelberg, Germany\\ 
  \email{grebel@ari.uni-heidelberg.de}
}

\date{Received ..... ; accepted .......}

 
\abstract{We determine the elemental abundances of nine red giant
  stars belonging to Palomar~14 (Pal~14). Pal~14 is an outer halo 
  globular cluster (GC) at a distance of $\sim 70$\,kpc. Our
  abundance analysis is based on high-resolution spectra and
  one-dimensional stellar model atmospheres. We derived the abundances
  for the iron peak elements Sc, V, Cr, Mn, Co, Ni, the
  $\alpha-$elements O, Mg, Si, Ca, Ti, the light odd
  element Na, and the neutron-capture elements Y, Zr, Ba,
    La, Ce, Nd, Eu, Dy, and Cu. Our data do not permit us to
      investigate light element (i.e., O to Mg) abundance variations. The
  neutron-capture elements show an r-process signature. We compare our
  measurements with the abundance ratios of inner and other outer halo
  GCs, halo field stars, GCs of recognized extragalactic origin, and stars in dwarf spheroidal galaxies (dSphs). 
  The abundance pattern of Pal~14 is almost
  identical to those of Pal~3 and Pal~4, the next distant members of
  the outer halo GC population after Pal~14. The abundance pattern of
  Pal~14 is also similar to those of the inner halo GCs, halo field
  stars, and GCs of recognized extragalactic origin, but differs from
  what is customarily found in dSphs field stars. The abundance properties of
    Pal~14 as well as those of the other outer halo GCs are thus compatible with
    an accretion origin from dSphs. Whether or not GC accretion played a role,
it seems that the formation conditions of outer halo GCs and GCs in dSphs were
similar.}


\maketitle


\section{Introduction}

Globular clusters (GCs), which are witnesses of the formation and
evolution of the Milky Way (MW), bear the traces of the environment in
which they formed.  GCs have the added advantage of being easily
identifiable objects whose distances can be measured relatively
easily and that are composed of stars of similar age and chemical
composition.  Halo GCs are of particular interest in this context, 
since they permit us to probe how the stellar halo of the MW formed. 
For example, \citet{searle78} examined a sample of
19 GCs in the inner and outer Galactic halo and proposed a hierarchial
halo formation scenario, in which the halo formed from accreted smaller
subunits. Observationally, there is considerable support for such a
scenario \citep{carolloetal07,carolloetal10,bell08,schlaufman09}.

Recent studies of GCs in the outer Galactic halo have concentrated on
deriving their chemical element pattern
\citep{koch10,kochetal09,leeetal06,cohen05a,cohen05b,shetroneetal01},
on testing gravitational theories
\citep{sollima10,baumgardt05,baumgardtetal09,haghi09,jordietal09}, on determining 
their structural and dynamical properties \citep{sollimaetal10,jordi10, hasanietal10}, and a better
understanding of the second parameter phenomenon. The latter
refers to additional parameters other than metallicity required to 
explain their horizontal branch (HB) morphology. Good candidates 
for these additional parameters include 
ages and star-to-star variations in the helium
content
\citep{dotteretal08,dotteretal10,grattonetal10,catelanetal01,stetsonetal99}.

Palomar~14 (Pal~14) is one of the most distant ($R_{\mathrm{GC}}$ =71$\pm$2
\,kpc), faint ($M_{V}$=$-$4.95$\pm$0.12\,mag), and diffuse ($r_{h}$= 46.1$\pm$2.9\,pc)
outer Galactic halo GCs \citep{sollimaetal10}. Color-magnitude diagram (CMD) studies 
of Pal~14 show that the cluster has a pair of degree-long tidal tails, a red
HB, a younger age ($\sim$12 Gyr) compared to inner halo cluster of
similar metallicities, and a metallicity ranging from $-$1.50 to
$-$1.60\,dex \citep{dacostaetal82,holland92,harris96,sarajedini97,hilker06,dotteretal08,jordietal09,jordi10,sollimaetal10}. The
cluster's first spectroscopic metallicity determination yielded
$-$1.60$\pm$0.18\,dex \citep{armandroffetal92} based on spectra of
the near-infrared Ca~II triplet region. The systemic velocity and the velocity
dispersion for Pal~14 were measured to be 72.28$\pm$0.12\,km~s$^{-1}$ 
and 0.38$\pm$0.12\,km~s$^{-1}$ by \citet{jordietal09}, respectively.

  A possible extragalactic origin of Pal~14 was discussed by several
  authors. \citet{lynden95} proposed that young GCs like Pal~14 have been
  accreted from dSphs. \citet{forbes10}, who examined the age$-$metallicity
  relation of several GCs, which include Pal~14 and are near to the
  Fornax--Leo--Sculptor great circle, suggested that these GCs were accreted
  by our Galaxy, except for three of them. \citet{sollimaetal10} found that the direction
  of the tidal tail of Pal~14 and its expected proper motion by \citet{lynden95} show an agreement, if
  Pal~14 is a part of the stream consisting of Fornax and Palomar~15. They
  also emphasized that Pal~14 has either an external orbit which is confined to the
  peripheral region of the Galactic halo or an extragalactic origin, based on the existing tidal tail of this cluster.

In this paper, our goal is to derive the elemental abundances of nine
stars belonging to Palomar 14 and to compare our measurements to those for
stars in inner and outer halo GCs, extragalactic GCs in dSphs, and the
halo field population as well
as dSph field stars. Our analysis is based on the spectra obtained by 
\citet{jordietal09}.  This is the first detailed abundance analysis of Pal~14
based on high-resolution spectroscopy.

\section{Observations}                                  

Our nine target stars were selected by \citet{jordietal09} from
among the brightest red giant members of Pal~14 using a CMD
obtained by \citet{hilker06}. The spectra of these stars, covering
the brightness range $V=17.37$--$19.41$, were obtained with
FLAMES/UVES in Service Mode in 2006 and 2007, with the original aim to
determine the velocity dispersion of Pal~14 \citep[for details,
  see][]{jordietal09}.  The spectra cover the wavelength ranges
$4760$--$5770$\,{\AA} and $5840$--$6840$\,{\AA}. UVES was used in
the RED~580\,nm setting (fiber diameter of 1\,arcsec). In this way, a
resolving power of $R \equiv$ $\lambda$/$\Delta{\lambda} \approx
{47,000}$ is reached. The data were reduced using the UVES pipeline
\citep[for details, see][]{jordietal09}.

After shifting the spectra to the rest frame, using the radial
velocities of \citet{jordietal09}, we rebinned all spectra by a factor
of three to increase their signal-to-noise ratios ($S/N$). Thus, we
  measured a FWHM of 0.02\,nm. The $S/N$ of each
spectrum around {$5500$ and} $6630$\,{\AA} --- the
regions have relatively low and high $S/N$ ratios for
all of spectra, respectively --- are listed in
Table~\ref{Kapsou}; for our stars the values are in the range
$6$--$36$ per pixel at the aforementioned wavelengths.

\begin{table}
  \caption[]{The properties of our target stars.}
  \label{Kapsou}
  \centering
  \begin{tabular}{r c c c c c c}
    \hline
    \noalign{\smallskip}
    ID$\mathrm{^a}$ & ID$\mathrm{^b}$ &  ${m_{v}}$&
    Exposure time& $v_{\mathrm{rad}}$ & $S/N$ $\mathrm{^c}$ &$S/N$ $\mathrm{^d}$ \\
    & & $\mathrm{[mag]}$&$\mathrm{[s]}$ &$\mathrm{[km~s^{-1}]}$& & \\
    \noalign{\smallskip}
    \hline
    \noalign{\smallskip}
    1 &       & 17.37 &  4$\times$3600 &72.53 &26&36 \\            
    2 & HV025 & 17.77 &  4$\times$3600 &72.76 &19&29 \\
    3 &       & 18.23 &  4$\times$3600 &71.75 &15&16 \\
    5 & HV007 & 18.52 &  4$\times$3600 &71.68 &10&11 \\
    6 & HH244 & 18.56 &  6$\times$3600 &72.58 &9 & 9 \\
    7 & HH201 & 18.70 &  6$\times$3600 &72.62 &6 &10 \\
    8 & HV043 & 18.84 &  6$\times$3600 &71.56 &6 &11 \\
    9 & HV104 & 19.05 &  6$\times$3600 &73.49 &6 &10 \\
    12& HV074 & 19.41 & 11$\times$3600 &71.83 &7 & 8 \\
    \noalign{\smallskip}
    \hline
  \end{tabular}
  \tablefoot{\\
    \tablefoottext{a}{Identification from \citet{hilker06}\\}
    \tablefoottext{b}{Identification from \citet{harris84,holland92}\\}
    \tablefoottext{c}{S/N ratio at $5500$\,{\AA} for all of spectra\\}
    \tablefoottext{d}{S/N ratio at $6630$\,{\AA} for all of spectra\\}}
\end{table}

\section{Stellar Parameters}

First, we derived the effective temperature $T_{\mathrm{eff}}$ and
surface gravity $\log g$ for each star from broad-band optical and
infrared photometry. To determine $T_{\mathrm{eff}}$, we used $VRI$ magnitudes
from \citet{sahaetal05} and $JHK$ magnitudes from the 2Micron All-Sky Survey
\citep[2MASS;][]{cutrietal03} and employed the \citet{alonsoetal99}
empirical color calibrations. We converted the magnitudes to the
Johnson system using the transformations of
\citet{bessell79,bessell83} and the Telescopio Carlos Sanchez (TCS)
system using the transformations of \citet{ramirez04}. For all stars,
we adopted a metallicity of $-1.5$\,dex (Harris 1996) and a reddening
of $E(B-V)=0.033$ \citep{dotteretal08}. Following the transformations,
we performed reddening corrections for each color, using the relations
given by \citet{besselletal98}. The effective temperatures of the
stars obtained by this procedure are listed in Table~\ref{atmospara}.
    
Surface gravities ($\log g$) were derived from photometry by means of
Yale-Yonsei isochrones \citep{yietal01,kimetal02}. For this, we used
10 and 11\,Gyr isochrones with an iron abundance of
$\mathrm{[Fe/H]=-1.5}$\,dex, $\alpha-$enhancement of
$\mathrm{[\alpha/Fe]=+0.3}$\,dex and the photometrically derived
$T_{\mathrm{eff}}$ of each star.

We determined the microturbulent velocity $\xi$ of each star by means
of the requirement that there shall be no trend of $\log \epsilon$
with equivalent width (EW) of Fe~I lines. We compared our results to the microturbulent velocities of
halo giants given by \citet{cayreletal04}, which have similar parameters as
the Pal~14 stars. The microturbulent velocities of the brightest three stars
that have relatively high $S/N$ ratios differ on average by
0.2~$\mathrm{km~s^{-1}}$ from $\xi$ derived by \citet{cayreletal04}. The other
stars that have lower $S/N$ ratios are on average higher by
0.4~$\mathrm{km~s^{-1}}$ than the results of \citet{cayreletal04}. The
differences may be due to the low $S/N$ ratio of our spectra and the abundance
errors of the strong lines \citep{kochetal09}. 

We note that the low $S/N$ in the majority of our spectra
prevents us from measuring weak lines ($<$25\,m{\AA}), which are a crucial
reference in determining an accurate microturbulence
as we are left only with the strong lines that show a high sensitivity to
this parameter. Furthermore, line asymmetries of weak lines in low $S/N$
spectra can also induce artificially increased values for $\xi$ \citep{magain84}. The results are listed in
Table~\ref{atmospara}.

\begin{table} 
  \caption[]{The stellar parameters of our target stars.}
  \label{atmospara}
  \centering
  \begin{tabular}{l c c c}
    \hline
    \noalign{\smallskip}
    ID &  $T_{\mathrm{eff}}$ $\mathrm{[K]}$ & $\log g$ $\mathrm{[cgs]}$ & $\xi$ $\mathrm{[km~s^{-1}]}$\\
    \noalign{\smallskip}
    \hline
    \noalign{\smallskip}
    1     & 4100 & 0.7 & 2.1\\            
    HV025 & 4160 & 0.8 & 2.3\\
    3     & 4600 & 1.6 & 2.3\\
    HV007 & 4520 & 1.4 & 2.4\\
    HH244 & 4400 & 1.2 & 2.3\\
    HH201 & 4440 & 1.3 & 2.5\\
    HV043 & 4630 & 1.6 & 2.1\\
    HV104 & 4660 & 1.7 & 2.6\\
    HV074 & 4730 & 1.8 & 2.3\\                       
    \noalign{\smallskip}
    \hline
  \end{tabular}
\end{table}

\begin{table*}
   \caption[]{Errors analysis for the star 1 and HV074.}
   \label{temlogmic}  
   \centering
   \begin{tabular}{l r r r r c r r r r r r l}
      \hline
      \noalign{\smallskip}
              &\multicolumn{6}{c}{1}&\multicolumn{6}{c}{HV074}\\
      \hline
      \noalign{\smallskip}      
      Species
      &\multicolumn{2}{c}{$\Delta
        T_{\mathrm{eff}}$}&\multicolumn{2}{c}{$\Delta\log g$}&\multicolumn{2}{c}{$\Delta \xi$}&\multicolumn{2}
      {c}{$\Delta T_{\mathrm{eff}}$}&\multicolumn{2}{c}{$\Delta\log g$}&\multicolumn{2}{c}{$\Delta \xi$}\\
      \noalign{\smallskip}
      &\multicolumn{2}{c}{$\mathrm{[K]}$}&\multicolumn{2}{c}{$\mathrm{[dex]}$}&\multicolumn{2}{c}{$\mathrm{[km~s^{-1}]}$}&
      \multicolumn{2}{c}{$\mathrm{[K]}$}&\multicolumn{2}{c}{$\mathrm{[dex]}$}&\multicolumn{2}{c}{$\mathrm{[km~s^{-1}]}$}\\
      \noalign{\smallskip}
        &$+$100&$-$100&$+$0.5&$-$0.5&$+$0.2&$-$0.2&$+$100&$-$100&$+$0.5&$-$0.5&$+$0.2&$-$0.2\\
      \noalign{\smallskip}
      \hline     
      \noalign{\smallskip}
 O~I &0.01   &$-$0.02&$-$0.21 &0.21   &$-$0.01&0.01  &$\cdots$&$\cdots$&$\cdots$&$\cdots$&$\cdots$&$\cdots$\\                  
Na~I &0.09   &$-$0.10&0.04    &$-$0.04&$-$0.03&0.02  &$\cdots$&$\cdots$&$\cdots$&$\cdots$&$\cdots$&$\cdots$\\                      
Mg~I &0.08   &$-$0.06&0.01    &$-$0.03&$-$0.07&0.07  &0.09    &$-$0.08 &0.06    &$-$0.06 &$-$0.05 &0.05    \\                 
Si~I &$-$0.04&0.07   &$-$0.12 &0.09   &$-$0.01&0.03  &0.03    &$-$0.02 &$-$0.04 &0.03    &$-$0.05 &0.04    \\           
Ca~I &0.14   &$-$0.14&0.02    &$-$0.03&$-$0.08&0.10  &0.11    &$-$0.12 &0.03    &$-$0.03 &$-$0.06 &0.06    \\                 
Sc~II&$-$0.03&0.02   &$-$0.19 &0.19   &$-$0.07&0.07  &$<$0.01 &0.01    &$-$0.21 &0.20    &$-$0.01 &0.02    \\                 
Ti~I &0.21   &$-$0.21&0.00    &$-$0.01&$-$0.07&0.08  &0.17    &$-$0.17 &0.02    &$-$0.04 &$-$0.05 &0.06    \\             
Ti~II&$-$0.02&0.03   &$-$0.19 &0.17   &$-$0.07&0.08  &$-$0.01 &0.01    &$-$0.20 &0.21    &$-$0.07 &0.07    \\                 
V~I  &0.23   &$-$0.22&$-$0.01 &0.00   &$-$0.04&0.06  &$\cdots$&$\cdots$&$\cdots$&$\cdots$&$\cdots$&$\cdots$\\                      
Cr~I &0.20   &$-$0.19&0.00    &$-$0.02&$-$0.12&0.15  &0.17    &$-$0.17 &0.02    &$-$0.04 &$-$0.12 &0.14    \\                 
Mn~I &0.11   &$-$0.10&$-$0.02 &0.00   &$-$0.05&0.06  &$\cdots$&$\cdots$&$\cdots$&$\cdots$&$\cdots$&$\cdots$\\                      
Fe~I &0.08   &$-$0.06&$-$0.07 &0.05   &$-$0.10&0.11  &0.14    &$-$0.14 &0.00    &$-$0.01 &$-$0.08 &0.08    \\                 
Fe~II&$-$0.13&0.15   &$-$0.27 &0.25   &$-$0.05&0.05  &$-$0.04 &0.05    &$-$0.22 &0.22    &$-$0.06 &0.06    \\                 
Co~I &0.09   &$-$0.07&$-$0.10 &0.06   &$-$0.03&0.04  &$\cdots$&$\cdots$&$\cdots$&$\cdots$&$\cdots$&$\cdots$\\                      
Ni~I &0.05   &$-$0.04&$-$0.10 &0.08   &$-$0.07&0.08  &0.12    &$-$0.12 &$-$0.03 &0.01    &$-$0.05 &0.05    \\                 
Cu~I &0.09   &$-$0.07&$-$0.09 &0.07   &$-$0.16&0.18  &0.14    &$-$0.14 &$-$0.02 &0.00    &$-$0.02 &0.02    \\                 
Y~II &$<$0.01&0.01   &$-$0.19 &0.17   &$-$0.05&0.07  &0.02    &$-$0.01 &$-$0.21 &0.20    &$-$0.03 &0.04    \\                
Zr~I &0.25   &$-$0.25&$-$0.02 &0.01   &$-$0.01&0.02  &$\cdots$&$\cdots$&$\cdots$&$\cdots$&$\cdots$&$\cdots$\\                     
Zr~II&$-$0.03&0.03   &$-$0.20 &0.18   &$-$0.02&0.03  &$\cdots$&$\cdots$&$\cdots$&$\cdots$&$\cdots$&$\cdots$\\                
Ba~II&0.04   &$-$0.03&$-$0.17 &0.18   &$-$0.16&0.17  &0.03    &$-$0.03 &$-$0.17 &0.20    &$-$0.14 &0.14    \\           
La~II&0.02   &$-$0.03&$-$0.20 &0.21   &$-$0.02&0.02  &$\cdots$&$\cdots$&$\cdots$&$\cdots$&$\cdots$&$\cdots$\\                      
Ce~II&0.01   &$-$0.01&$-$0.19 &0.18   &$-$0.03&0.04  &$\cdots$&$\cdots$&$\cdots$&$\cdots$&$\cdots$&$\cdots$\\                
Nd~II&0.02   &$-$0.03&$-$0.19 &0.17   &$-$0.11&0.13  &0.02    &$-$0.03 &$-$0.20 &0.21    &$-$0.05 &0.05    \\                 
Eu~II&$-$0.03&0.02   &$-$0.21 &0.21   &$-$0.03&0.03  &$\cdots$&$\cdots$&$\cdots$&$\cdots$&$\cdots$&$\cdots$\\                
Dy~II&0.03   &$-$0.03&$-$0.20 &0.18   &$-$0.02&0.03  &$\cdots$&$\cdots$&$\cdots$&$\cdots$&$\cdots$&$\cdots$\\                      
\noalign{\smallskip}
\hline
\end{tabular}
\end{table*}

\section{Abundance Analysis}

\subsection{Model Atmospheres}

We employed a Linux version \citep{sbordoneetal04} of the ATLAS9
code \citep{kurucz93a,kurucz05,sbordoneetal04} for computing
individual model atmospheres for the stars of our sample. In the
computation, local thermodynamic equilibrium (LTE), plane-parallel
geometry, hydrostatic equilibrium, and no convective overshooting
are assumed.

We computed tailored models for each of our stars, starting from grids of
ATLAS9 model atmospheres that were computed by \citet{castelli03} for a
metallicity of $\mathrm{[M/H]=-1.5}$\,dex with $\alpha$-enhanced new opacity
distribution functions (ODFs).
 
\begin{figure}[ht]
  \centering
  \includegraphics[bb=45 210 543 791,width=9cm,clip]{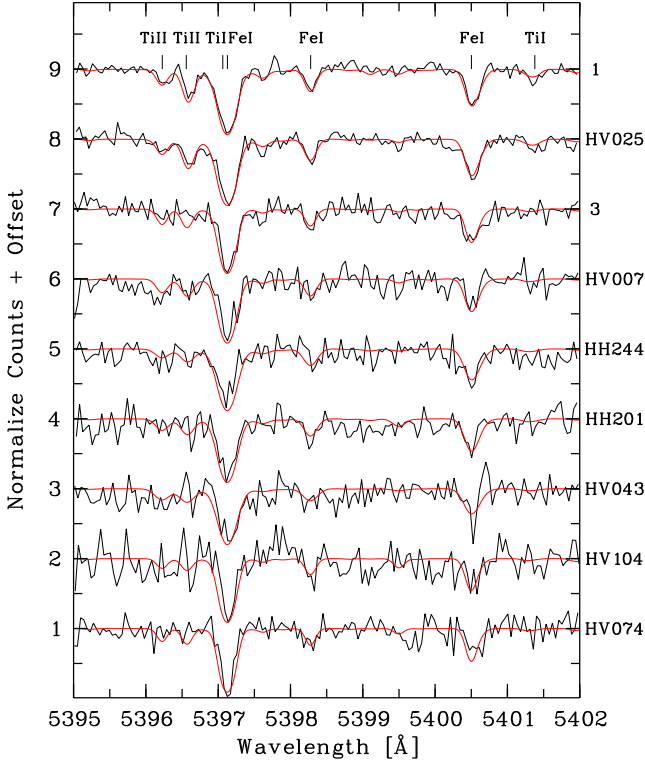}
  \caption{The $5395$-$5402$\,{\AA} region of the observed (black
    lines) and synthetic (red lines) spectra of all nine target
    stars.}
  \label{aasynl}
\end{figure}

\begin{figure}[ht]
  \centering
  \includegraphics[bb=45 210 543 791,width=9cm,clip]{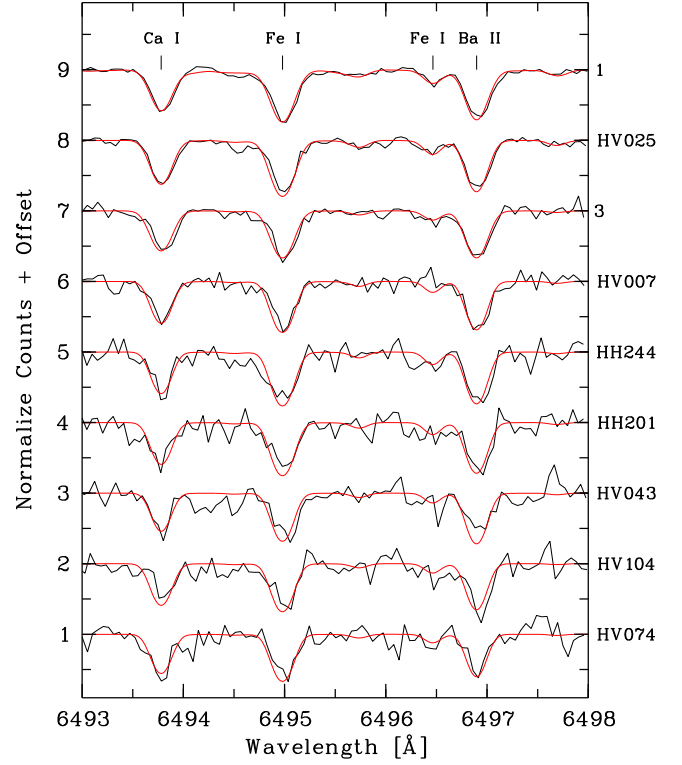}
  \caption{The $6493$-$6498$\,{\AA} region of the observed (black
    lines) and synthetic (red lines) spectra of all nine target
    stars.}
  \label{aasyn}
\end{figure}

\subsection{Line list and equivalent width measurement}

We compiled a line list from \citet{kochetal09,koch10}, and
the Vienna Atomic Database
\citep[VALD,][]{piskunovetal95,kupkaetal99,ryabchikovaetal99}. For Fe~II
lines, we used the $gf$ values determined by \citet{melendez09}. Line
lists including hyperfine structure (HFS) splitting for Y~II, Ba~II,
La~II, and Eu~II were downloaded from Kurucz's web page.
   
Equivalent width measurements were done by fitting a Gaussian profile to the lines
simultaneously with a straight-line continuum, where the continuum and
line regions were chosen interactively. To estimate the uncertainties in the EW
measurements, we used the formula of \citet{cayrel88}, which estimates the
uncertainty of EW measurements depending on the full width at half maximum of
the line, the pixel size, and the $S/N$ of spectrum. For the abundance
analysis, we did not use lines with EWs larger than
  186\,m{\AA} and uncertainties in the EW
measurements larger than 30$\%$. This
uncertainty in the EW measurement leads to an error of $<$0.16\,dex in the derived abundance. From these measurements,
abundances were derived using the WIDTH9 code \citep{kurucz05,sbordoneetal04}, which uses ATLAS9 model atmospheres
to compute line formation in LTE.

The line list, measured EWs, and the abundances calculated
for each line are provided in Table A.1 and A.2.
  
\begin{table} 
  \caption[]{The mean abundances of the species observed in Palomar~14}
  \label{aburesu}
  \centering
  \tiny
  \begin{tabular}{l r c c c}
    \hline
    \noalign{\smallskip}
    Element & $\mathrm{[X/Fe]}$ & $\sigma_{s}$ $\mathrm{^a}$&$\sigma_{tot}$ & $\log\epsilon_{\odot}$ $\mathrm{^b}$\\
    \noalign{\smallskip}
    \hline                         
    \noalign{\smallskip}           
    $\mathrm{[FeI/H]}$    &$-$1.44&0.03&0.16&7.50\\
    $\mathrm{[FeII/H]}$   &$-$1.23&0.05&0.33&7.50\\
    O I$\mathrm{^{syn}}$   &0.59   &0.09&0.41&8.69\\
    Na~I                  &$-$0.14&0.09&0.23&6.24\\
    Mg~I                  &0.37   &0.08&0.31&7.60\\         
    Si~I                  &0.42   &0.10&0.23&7.51\\  
    Ca~I                  &0.29   &0.06&0.24&6.34\\                                         
    Sc~II                 &0.12   &0.09&0.41&3.15\\                                     
    Ti~I                  &0.30   &0.10&0.26&4.95\\                                         
    Ti~II                 &0.18   &0.07&0.43&4.95\\                                       
    V~I                   &0.20   &0.07&0.28&3.93\\                                      
    Cr~I                  &$-$0.01&0.07&0.30&5.64\\                                    
    Mn~I$\mathrm{^{hfs}}$  &$-$0.27&0.07&0.20&5.43\\
    Co~I                  &0.02   &0.06&0.20&4.99\\                                      
    Ni~I                  &0.04   &0.07&0.24&6.22\\                                          
    Cu~I$\mathrm{^{hfs}}$  &$-$0.35&0.04&0.25&4.19\\
    Y~II                  &$-$0.13&0.11&0.49&2.21\\
    Zr~I                  &0.08   &0.04&0.29&2.58\\
    Zr~II                 &0.03   &0.09&0.40&2.58\\ 
    Ba~II                 &$-$0.11&0.08&0.48&2.18\\
    La~II                 &0.44   &0.23&0.40&1.10\\
    Ce~II                 &0.29   &0.15&0.46&1.58\\ 
    Nd~II                 &0.50   &0.08&0.46&1.42\\ 
    Eu~II$\mathrm{^{hfs}}$ &0.56   &0.11&0.51&0.52\\ 
    Dy~II                 &0.58   &0.16&0.43&1.10\\                         
    \noalign{\smallskip}                 
    \hline                               
  \end{tabular}
  \tablefoot{\\
  \tablefoottext{a}{The standard deviations of the abundance ratios over the square root of the 
    number of stars\\}
  \tablefoottext{b}{Solar abundance ratios are taken from \cite{asplundetal09}\\}}
\end{table}

\subsection{Spectrum synthesis}

Since the EW method is not appropriate for blended lines and lines where
hyperfine structure (HFS) splitting has to be taken into account, we
applied the spectrum synthesis method for HFS and blended lines such
as the $6300$\,{\AA} forbidden O line: we produced synthetic spectra
with the SYNTHE code \citep{kurucz93b,kurucz05} and then those spectra
were convolved with a Gaussian profile that includes the broadening
effects due to the instrumental profile, and the macroturbulence
velocity. The abundances of the species were adjusted until the
observed and synthetic spectrum were in good agrement.  We accounted
for the effect of HFS for Mn I, Cu I, and Eu II, but ignored the
effect as negligible (abundance difference $<$0.02\,dex) for V I, Sc
II, Co I, Y II, Ba II, and La II. In Figure~\ref{aasynl} and Figure~\ref{aasyn}, we show
samples of the synthetic and observed spectra in the
$5395-5402$\,{\AA} and $6493-6498$\,{\AA} regions, respectively.

\begin{figure*}[ht]
  \centering
  \includegraphics[bb=29 515 615 735,width=15.5cm,clip]{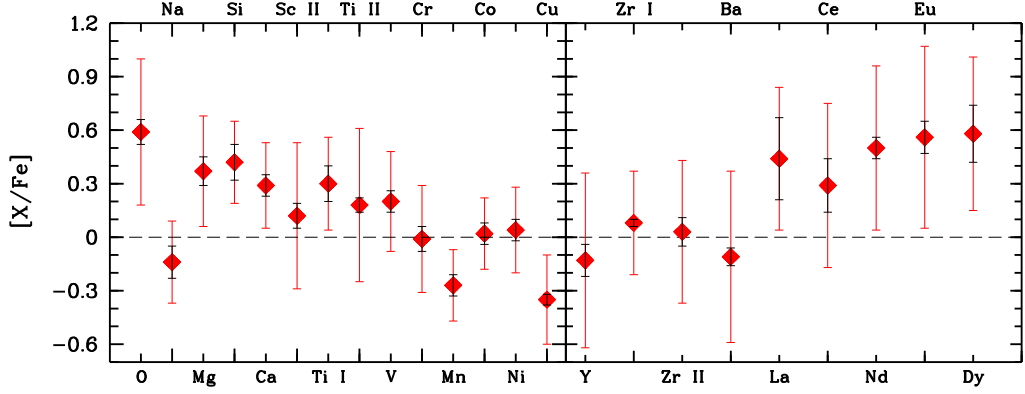}
  \caption{Mean abundance ratios of the Pal~14 giants of our
    sample. The black error bars indicate the standard deviations of the
    abundance ratios over the square root of the 
    number of stars, ($\mathrm{\sigma_{s}=\sigma_{s}/\sqrt{N}}$). The red
      error bars indicate the total
      errors, which involve the random and systematic uncertainties of the abundance ratios.}
  \label{aburatioall}
\end{figure*}

The mean of the abundances determined with the EW measurements and
spectrum synthesis for all our stars is listed in
Table~\ref{aburesu}. These abundance ratios, together with their
  uncertainties, of our all stars
  are shown in Figure~\ref{aburatioall}. The abundance ratios of the neutral and ionized species are given relative
to Fe~I and Fe~II, respectively, except for the O abundance ratio.

\subsection{Abundance errors}

The random errors, $\mathrm{\sigma=\sigma/\sqrt{N}}$ ($N$ is the number of lines for each element), of the
abundance ratios were determined from the uncertainties in the measurements of
the EW, which were estimated using the relation of
\citet{cayrel88} for each line. The random errors in the elemental abundances, which are measured 
with spectrum synthesis, were given by the standard deviation of individual
measurements over the square root of number of lines. We adopted an uncertainty of $0.10$\,dex for species 
that were measured with spectrum synthesis and also with only one detected line. 
The random errors are listed in Table~\ref{individui}.

The systematic errors were estimated from the uncertainties in the stellar
parameters for each element. For each model atmosphere used in our analysis,
the stellar parameters $T_{\mathrm{eff}}$, $\log g$, $\xi$ were varied within
an uncertainty of $\pm100$\,K, $\pm0.5$\,dex, and $\pm0.2$\,$\mathrm{km~s^{-1}}$,
respectively. As an example, the deviations of the abundances, which were
obtained from the unchanged atmosphere parameters, of the brightest and faintest
stars of Pal~14 are listed in Table~\ref{temlogmic}. The largest uncertainty in the abundance ratios is due to
$T_{\mathrm{eff}}$. The abundance ratios of ionized and neutral species vary with
0.2\,dex and 0.1\,dex, respectively, when the value of $\log g$ is changed by
an amount of 0.5\,dex.
  
The total errors on our abundance ratios were calculated as
$\mathrm{\sigma_{tot}=\sqrt{\sigma^2+\sigma_{sys}^2}}$ for each target star and
are listed in Table~\ref{aburesu}, as $\mathrm{\sigma_{tot}=\sigma_{tot}/\sqrt{N}}$ ($N$ is the number of stars).

\section{Abundance results}
   
\subsection {Oxygen and sodium}

Oxygen abundances were derived from the $6300$ and $6363$\,{\AA} lines
in the two brightest stars, while only the $6300$\,{\AA} line was
detected in the spectra of next three brighest stars.  We determined
the oxygen abundances with the spectrum synthesis method, since the
oxygen lines in the spectra are quite weak and affected by the blends. The $S/N$ of the spectra of the four remaining stars
is too low for oxygen abundance measurements. The mean of the oxygen
abundance ratio of the five brightest stars in our sample is
$\mathrm{[O/Fe]}=+0.59\pm0.09$\,dex. We note that throughout this paper, we
use the solar abundances of \citet{asplundetal09} when computing
abundance ratios relative to the Sun.

Sodium abundances were determined for six of our target stars from EW
measurements using the $5682$ and $5688$\,{\AA} lines. We did not take into account non-LTE effects for these
  lines, since the 5682 and 5688\,{\AA} lines do not have any significant non-LTE effect
\citep{takedaetal03}. We find a mean Na abundance ratio of
$\mathrm{[Na/Fe]=-0.14\pm0.09}$\,dex for Pal~14.
  
\subsection{Magnesium to titanium}

We used the $5528$ and $5711$\,{\AA} neutral lines to derive the magnesium
abundance of each star. These lines were detected in the spectra of
all nine stars. No non-LTE corrections were applied, because according
to \citet{gehrenetal04} they are neglible compared to the
uncertainties of our Mg abundances. The Mg abundance ratios are
comparable to each other for seven of the target stars, while there are
significant star-to-star variations between the star 3 and HH201.

Silicon abundances were determined using the $5690$, $5948$, and
$6155$\,{\AA} neutral lines for six of the stars. Calcium and titanium
abundances were derived for all target stars. The difference
between neutral and ionized titanium is
$\mathrm{[Ti~I/Ti~II]=0.12\pm0.12}$\,dex. 

\begin{figure}[ht]
  \centering
  \includegraphics[bb=37 439 553 751,width=9cm,clip]{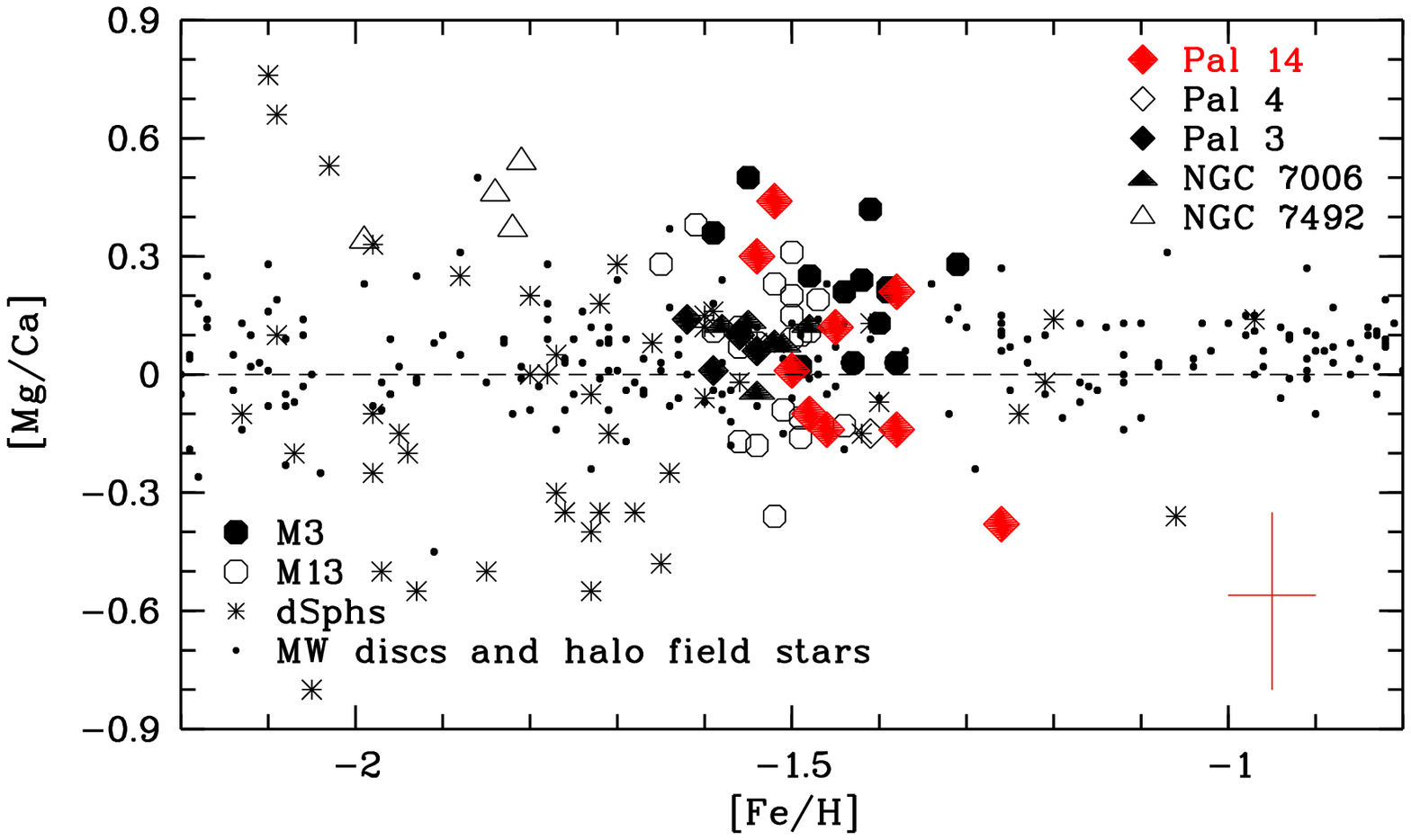}
  \caption{$\mathrm{[Mg/Ca]}$ versus [Fe/H]. The
      Pal~4 and Pal~3 are taken from \citet{koch10,kochetal09}.The NGC~7006
      data are from \citet{kraftetal98}, the NGC~7492 data from
    \citet{cohen05b}, the M3 data from \citet{cohen05a}, the M13 data from
    \citet{cohen05a}, dSph star data from
    \citet{vennetal04,kochetal08a,shetroneetal09,cohen09,cohen10}, and the MW disc
   and halo field star data from
   \citet{johnson02,cayreletal04,vennetal04,sobecketal06,ishigakietal10}. Corrections were applied to
    the literature data when Solar abundances differing from those of
    \citet{asplundetal09} were used in computing abundance ratios. The
      red error bar indicates the average random error in the abundance ratios
      of our target stars}.
  \label{mgca}
\end{figure}

Furthermore, the $\mathrm{[Mg/Ca]}$ ratios of Pal~14 stars are
noteworthy. O and Mg are products of hydrostatic nuclear burning,
while Si, Ca and Ti are mostly produced as a result of explosive
nuclear burning in SN~II. Therefore, the $\mathrm{[Mg/Ca]}$ ratio is
an indicator of the progenitor mass \citep{koch10a,koch10}. In Figure
~\ref{mgca}, we show the $\mathrm{[Mg/Ca]}$ ratios of the Pal~14 stars
as well as the comparison clusters, Galactic halo field and dSph
stars. Pal~14 exhibits a larger scatter in comparison to the other
outer halo GCs, while the dispersion is similar to that in M13. On
the other hand, star~3 of our sample, having the lowest
$\mathrm{[Mg/Ca]}$ ratio, differs from the other stars of Pal~14 as
well as from halo stars.  According to the study of \citet{heger10},
$\mathrm{[Mg/Ca]}\approx-0.38$\,dex can be produced by SN~II with a
progenitor mass of $15$\,M$_{\odot}$, while the highest
$\mathrm{[Mg/Ca]}$ ratio occurs in SN~II with a progenitor mass of
$23$\,M$_{\odot}$. This suggests that a range of SN~II of different
masses contributed to the chemical inventory of Pal~14.  The Mg, Ca,
Si and Ti abundance ratios of the individual stars are listed in
Table~\ref{individui}.

\subsection{Iron peak elements}

We derived mean values of $\mathrm{[Fe~I/H]=-1.44\pm0.03}$ and
$\mathrm{[Fe~II/H]=-1.23\pm0.05}$ for the nine stars of our Pal~14
sample; i.e., the average difference between the Fe abundance derived
from Fe~I and Fe~II lines is $\log\epsilon(\mathrm{Fe~I})-\log\epsilon(\mathrm{Fe~II})=-0.21\pm0.06$. 
One reason for the imbalance is most likely that ionization equilibrium is not
fulfilled in our analysis. That is, $\log g$ was not derived by forcing
neutral and ionized iron to be the same. Note that an decrease of $\log g$ by
0.5$\pm0.2$\,dex would establish ionization equilibrium. In this case, the abundances
of Fe~I and Fe~II would be higher by 0.02$\pm0.01$ and 0.2$\pm0.1$\,dex, respectively.

Our mean iron abundance agrees well with the metallicity of
$-$1.60$\pm$0.18\,dex derived from spectra
covering the Ca~II triplet region by \citet{armandroffetal92} 
and photometric $\mathrm{[Fe/H]}$ estimates ranging from $-$1.50 to $-$1.60\,dex based on CMDs of Pal~14
\citep{dacostaetal82,holland92,harris96,sarajedini97,hilker06,dotteretal08}.

We also measured the abundances of the iron peak elements Sc, V, Cr,
Mn, Co, and Ni. The Cr abundance is slightly underabundant by
$-0.01$\,dex, the Ni, Sc, and Co
abundances are overabundant by $0.04$, $0.12$, and
  $0.02$\,dex, respectively, while the Mn abundance is
underabundant by $-0.27$\,dex with respect to the solar values. The V abundance is 
overabundant by $0.20$\,dex as compared to solar values. 

\begin{figure}[ht]
  \centering
  \includegraphics[bb=37 439 553 751,width=9cm,clip]{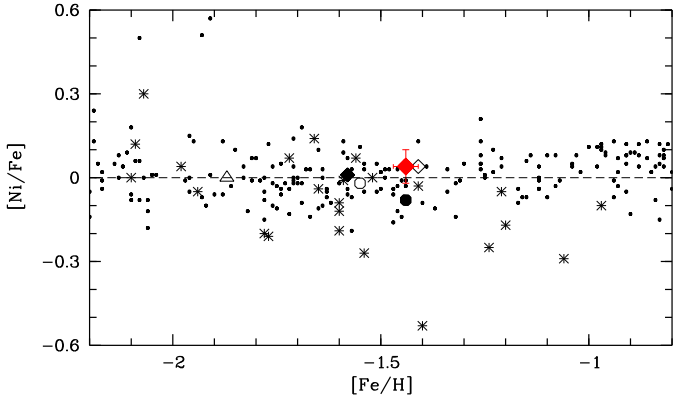}
  \caption{$\mathrm{[Ni/Fe]}$ versus $\mathrm{[Fe/H]}$. The symbols
      and references are the same as in Figure~\ref{mgca}.}
  \label{niratio}
\end{figure}

\begin{figure}[ht]
  \centering
  \includegraphics[bb=37 439 553 751,width=9cm,clip]{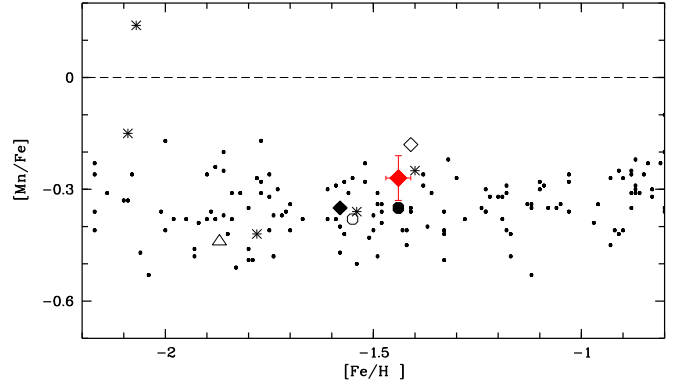}
  \caption{$\mathrm{[Mn/Fe]}$ versus $\mathrm{[Fe/H]}$. The symbols
    and references are the same as in Figure~\ref{mgca}.}
  \label{mnratio}
\end{figure}

As shown in Figure~\ref{niratio}, the Ni abundance ratio is $\sim 0$
over the entire metallicity range, but the scatter increases with
decreasing $\mathrm{[Fe/H]}$. Moreover, the Ni abundance ratio
  increases to higher metallicities. Pal~14, with $\mathrm{[Ni/Fe]}=0.04\pm0.07$, agrees with this trend. Its Mn
abundance ratio is slightly above the Galactic abundance ratio trend, but
it agrees with both the other Galactic GCs and dwarf spheroidal galaxies
(dSphs) of similiar metallicity, as shown in Figure ~\ref{mnratio}. The iron and iron peak element
abundances of individual stars are given in Table~\ref{individui}.

\begin{figure}[ht]
  \centering
  \includegraphics[bb=37 439 553 751,width=9cm,clip]{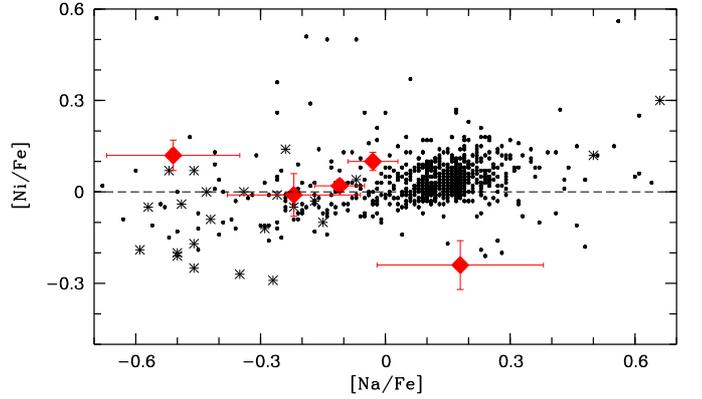}
  \caption{$\mathrm{[Ni/Fe]}$ versus $\mathrm{[Na/Fe]}$. The symbols
    and references are the same as in Figure~\ref{mgca}.}
  \label{naniratio}
\end{figure}

The production of Ni is governed by the neutron excess, which depends
on the amount of $^{23}$Na produced earlier during hydrostatic C burning,
resulting from SN~II events. Thus, the presence of a Na-Ni correlation
indicates a dominance of enrichment by SN~II. The Na-Ni relation is
modified by contributions from SN~Ia, since here the Ni production is
independent of the presence of Na \citep{vennetal04,letarteetal10}.
As can be seen in Figure~\ref{naniratio}, we do not detect a
significant correlation in our sample of Pal~14 stars.

\subsection {Neutron-capture elements}

We measured the abundances of the neutron-capture elements Y, Zr, Ba,
La, Ce, Nd, Eu and Dy. Lines of Zr are only detected in the two
brightest stars of our sample.  Y and Ba are notably underabundant
with respect to the abundance ratio in the Sun, while the other
neutron-capture elements are slightly overabundant. The abundance
ratios of the individual stars are listed in Table~\ref{individui}.

\section{Discussion}

\subsection{$\alpha-$elements}

We find a mean $\alpha$-enhancement of $0.34\pm0.17$\,dex for Pal 14, where
the $\alpha-$element abundance is calculated as $\mathrm{(Mg+Ca+Ti)/3}$ \citep{vennetal04,
kochetal08b}.  Mg, Ca, and Ti are all even-Z elements 
expected to have been produced during shell-burning in short-lived 
massive stars and to have been ejected in SN II events. 

\begin{figure}[ht]
  \centering
  \includegraphics[bb=37 439 553 751,width=9cm,clip]{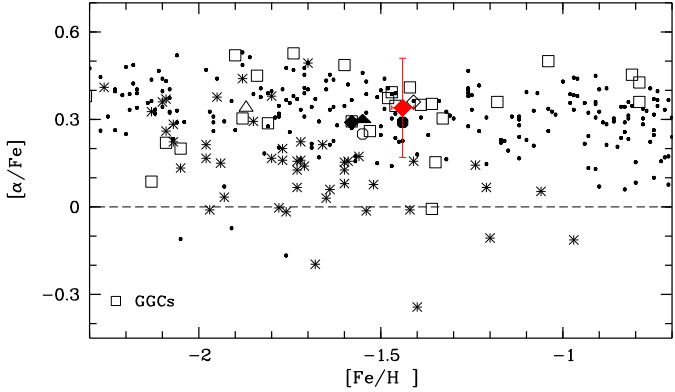}
  \caption{The mean $\mathrm{[\alpha/Fe]}$ abundance ratio.The symbols
    and references are the same as in Figure~\ref{mgca}. The data for
      Galactic globular clusters are
    taken from \citet{pritzletal05}.}
  \label{alpha}
\end{figure}

In Figure~\ref{alpha}, we compare the mean [$\alpha$/Fe] abundance ratio of
Pal~14 with the [$\alpha$/Fe] values of Galactic halo field stars
\citep{cayreletal04, sobecketal06, ishigakietal10, johnson02, vennetal04}, the
inner and outer halo GCs \citep{koch10, kochetal09, cohen05b, cohen05a, pritzletal05, kraftetal98}, and
field stars in the dSphs Carina, Draco, Hercules,
Fornax, Leo~II, Sculptor and Ursa Minor
\citep{vennetal04,kochetal08a,shetroneetal09,cohen09,cohen10,adenetal11}. The [$\alpha$/Fe] ratio of Pal~14 turns out to
be similar to the Galactic halo field stars and GCs, but differs from that of
field stars in typical dSphs. This suggests that
Pal~14 likely formed in an environment different from that of dSph field
stars, which experienced a low star formation rate (SFR), so that the
contribution of iron from supernovae type Ia (SN~Ia) to the chemical
enrichment sets in at lower metallicity compared to environments with
high SFR. For a comparison with GCs from dSphs, see Section~\ref{sec:extragalacticgcs}.

\subsection {Abundance variations of the light elements}

Correlations or anticorrelations among the light elements (such as O,
Na, Mg, Al) have been observed in many GCs \citep[e.g.,][]{harbeck03,
  kayser08, carrettaetal09a, carrettaetal09b, grattonetal04}. Two
main scenarios have been proposed to explain this behaviour: (1)
evolutionary mixing, and (2) primordial scenarios. According to the
evolutionary scenario, the light elements are produced in the deeper
and hotter layers of stars and transported to the surface by
non-standard mixing processes \citep{grattonetal04}. In the primordial scenario, which is based on either
classical self-enrichment (three generations of stars) or pre-enrichment scenarios (two
generations of stars), at least two generations of stars have contributed to
the enrichment of the GC \citep{grattonetal04,prantzos06}. 

According to the study of 19 GCs of \citet{carrettaetal09a,carrettaetal09b}, 
two generation of stars are responsible for the trend seen in
these light elements. According to the same study, the first generation
stars (notably massive and/or intermediate-mass asymptotic giant branch (AGB)
stars) form the so-called primordial (P) component, which in terms of
their O and Na content are similiar to field stars of the same
$\mathrm{[Fe/H]}$ and have high-O and low-Na abundances. Their
abundance ratios are consistent with supernova nucleosynthesis.
The P component is present in all clusters. The rest of the stars are
associated with the second generation stars (formed by material of the first
generation polluters), defined as I (intermediate) and E (extreme) components
with respect to the ratios $\mathrm{[O/Na]>-0.9}$\,dex and $\mathrm{[O/Na]<-0.9}$\,dex, respectively. 

\begin{figure}[ht]
  \centering
  \includegraphics[bb=37 439 553 751,width=9cm,clip]{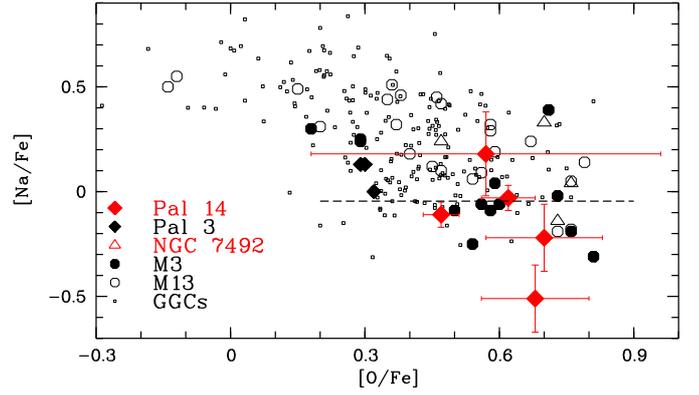}
  \caption{The Na-O anticorrelation for five red giants of Pal~14. The
      references are the same as in
      Figure~\ref{mgca}.The data for Galactic globular cluster stars are taken from \citet{carrettaetal09a}.}
  \label{naoanticor}
\end{figure}

Considering these scenarios, we probe the existence of the Na--O
anticorrelation among Pal~14 stars. According to the relation given by
\citet{carrettaetal09a}, star~1, 3, and HH244, which have low Na and
high O abundance ratios may be related to the primordial component (below the
dashed horizontal line in Figure ~\ref{naoanticor}), while
the rest of the stars are likely second generation stars (above the dashed
line in Figure~\ref{naoanticor}). 

In order to be able to investigate this
possibility, we also need to investigate the Mg--Al
anticorelation. This is useful because the first and second generations do
not only lead to variations of the Na and O abundance but also to variations
of the Al and Mg abundances in the same proton-capture reactions. Star
3 has a lower Mg abundance compared to the other stars. 
If the Al abundance of this star had been measured and
were lower compared to the other ones, our conclusion would be
confirmed. Unfortunately, we have neither detected the Al line nor
measured the O and Na abundances for all of our stars, thus there is not
enough evidence to draw such a conclusion. 
 
We also carried out the Spearman's rank test in order to understand the existence
of the possible anti-correlation between Na and O in our five stars. 
We found a correlation coefficient, $\rho$, of $-$0.6. When we compared our calculated
$\rho$ with the critical values of the coefficient at the 5$\%$ level of
significance for various numbers of pairs, we found that any relation between
Na and O in our stars has a probability of less than 95$\%$. However, given
the small number of stars we cannot draw any firm conclusion.

On the other hand, it is noteworthy that the O abundance ratio of our
target stars shows a rather small scatter and is fairly high for all 
five stars. In contrast, the Na abundances show a much larger range
of values, although none of the stars shows a particularly high Na
abundance. Thus, most of the analyzed stars in Pal~14 likely belong
to the primordial population, since otherwise we would expect a
different behavior, namely high Na and low O abundances. But again, the large
uncertainties in Na and O abundance ratios and the small number of our
target stars do not allow us to draw this kind of conclusion.

\begin{table*}
   \caption[]{The abundance ratios of the nine Pal~14 giants.}
   \label{individui}  
   \centering
   \tiny
   \begin{tabular}{l r c r r c r r c r r c l r c r}
      \hline
      \noalign{\smallskip}
      ID of stars & $\mathrm{[Fe~I/H]}$ &$\sigma$ &N& $\mathrm{[Fe~II/H]}$&
      $\sigma$&N&$\mathrm{[O/Fe]}$*&$\sigma$&N& $\mathrm{[Na/Fe]}$&$\sigma$&N&$\mathrm{[Mg/Fe]}$&$\sigma$ &N\\
      \noalign{\smallskip}
      \hline     
       \noalign{\smallskip}
         1   &$-$1.52&0.01&60&$-$1.31&0.04&6&0.47    &0.04    &2       &$-$0.11 &0.06    & 2      &   0.63&0.05&2\\
       HV025 &$-$1.50&0.01&59&$-$1.39&0.04&6&0.62    &0.06    &2       &$-$0.03 &0.06    & 2      &   0.25&0.02&2\\
         3   &$-$1.26&0.02&49&$-$1.13&0.07&5&0.68    &0.12    &1       &$-$0.51 &0.16    & 1      &$-$0.04&0.08&2\\
       HV007 &$-$1.48&0.04&31&$-$1.18&0.38&1&0.57    &0.39    &1       &   0.18 &0.20    & 1      &   0.35&0.06&2\\
       HH244 &$-$1.45&0.04&26&$-$1.24&0.08&5&0.70    &0.13    &1       &$-$0.22 &0.16    & 1      &   0.34&0.09&2\\
       HH201 &$-$1.38&0.04&29&$-$0.89&0.13&3&$\cdots$&$\cdots$&$\cdots$&$\cdots$&$\cdots$&$\cdots$&   0.04&0.15&2\\
       HV043 &$-$1.54&0.08&14&$-$1.36&0.13&2&$\cdots$&$\cdots$&$\cdots$&$\cdots$&$\cdots$&$\cdots$&   0.73&0.34&2\\
       HV104 &$-$1.38&0.05&18&$-$1.38&0.13&2&$\cdots$&$\cdots$&$\cdots$&$\cdots$&$\cdots$&$\cdots$&   0.66&0.60&2\\
       HV074 &$-$1.46&0.08& 8&$-$1.18&0.18&2&$\cdots$&$\cdots$&$\cdots$&$\cdots$&$\cdots$&$\cdots$&   0.43&0.22&2\\
      \noalign{\smallskip}
      \hline
      \hline
      \noalign{\smallskip}
      ID of stars & $\mathrm{[Si/Fe]}$ &$\sigma$ &N&
      $\mathrm{[Ca/Fe]}$&$\sigma$ &N&$\mathrm{[Sc/Fe]}$&$\sigma$&N&
      $\mathrm{[Ti~I/Fe]}$&$\sigma$&N&$\mathrm{[Ti~II/Fe]}$&$\sigma$ &N\\
      \noalign{\smallskip}
      \hline     
       \noalign{\smallskip}
         1   &0.20    &0.05    &3       &0.19&0.02&13&0.11    &0.05    &6       &0.27&0.02&24&   0.08&0.06&7 \\
       HV025 &0.29    &0.06    &3       &0.24&0.03&13&0.11    &0.05    &6       &0.30&0.02&22&   0.27&0.06&7 \\
         3   &0.27    &0.12    &1       &0.34&0.05&12&0.19    &0.09    &6       &0.29&0.05&19&   0.19&0.11&5 \\
       HV007 &0.39    &0.13    &2       &0.45&0.09&9 &0.12    &0.40    &4       &0.55&0.09&6 &   0.20&0.39&5 \\
       HH244 &$\cdots$&$\cdots$&$\cdots$&0.22&0.10&6 &0.16    &0.12    &4       &0.10&0.07&8 &   0.26&0.12&5 \\
       HH201 &0.55    &0.20    &1       &0.18&0.11&8 &$-$0.17 &0.18    &2       &0.12&0.09&6 &$-$0.20&0.21&2 \\
       HV043 &$\cdots$&$\cdots$&$\cdots$&0.43&0.22&3 &0.50    &0.20    &2       &1.01&0.11&5 &   0.27&0.14&4 \\
       HV104 &$\cdots$&$\cdots$&$\cdots$&0.45&0.10&7 &$\cdots$&$\cdots$&$\cdots$&0.79&0.14&4 &   0.14&0.15&2 \\
       HV074 &0.77    &0.17    &2       &0.57&0.12&7 &$-$0.15 &0.25    &2       &0.34&0.19&7 &   0.24&0.06&3 \\
      \noalign{\smallskip}
      \hline
      \hline
      \noalign{\smallskip}
      ID of stars & $\mathrm{[V/Fe]}$ &$\sigma$ &N& $\mathrm{[Cr/Fe]}$&$\sigma$
      &N&$\mathrm{[Mn/Fe]}$*&$\sigma$&N& $\mathrm{[Co/Fe]}$&$\sigma$ &N&$\mathrm{[Ni/Fe]}$&$\sigma$ &N\\
      \noalign{\smallskip}
      \hline     
       \noalign{\smallskip}
         1   &0.11    &0.02    &15      &   0.02 &0.05    &8       &$-$0.27 &0.03    &3       &0.16    &0.04    &4       &   0.02&0.02&18 \\
       HV025 &0.18    &0.02    &15      &$-$0.03 &0.05    &7       &$-$0.39 &0.04    &3       &0.02    &0.05    &4       &   0.10&0.03&18 \\
         3   &0.30    &0.05    &5       &   0.15 &0.11    &7       &$-$0.33 &0.05    &3       &$\cdots$&$\cdots$&$\cdots$&   0.12&0.05&11 \\
       HV007 &0.24    &0.07    &6       &   0.10 &0.13    &4       &$-$0.08 &0.07    &2       &$\cdots$&$\cdots$&$\cdots$&$-$0.24&0.08&6  \\
       HH244 &0.18    &0.07    &6       &$-$0.08 &0.10    &6       &$-$0.29 &0.08    &2       &$\cdots$&$\cdots$&$\cdots$&$-$0.01&0.07&11 \\
       HH201 &0.11    &0.10    &3       &$-$0.07 &0.11    &6       &$\cdots$&$\cdots$&$\cdots$&$\cdots$&$\cdots$&$\cdots$&   0.05&0.11&5  \\
       HV043 &$\cdots$&$\cdots$&$\cdots$&$\cdots$&$\cdots$&$\cdots$&$\cdots$&$\cdots$&$\cdots$&$\cdots$&$\cdots$&$\cdots$&   0.25&0.15&3  \\
       HV104 &$\cdots$&$\cdots$&$\cdots$&$-$0.39 &0.22    &2       &$\cdots$&$\cdots$&$\cdots$&$\cdots$&$\cdots$&$\cdots$&   0.22&0.10&3  \\
       HV074 &$\cdots$&$\cdots$&$\cdots$&   0.13 &0.27    &2       &$\cdots$&$\cdots$&$\cdots$&$\cdots$&$\cdots$&$\cdots$&$-$0.12&0.14&3  \\
      \noalign{\smallskip}
      \hline
      \hline            
       \noalign{\smallskip}            
       ID of stars & $\mathrm{[Cu/Fe]}$*&$\sigma$ &N& $\mathrm{[Y/Fe]}$&
       $\sigma$ &N&$\mathrm{[Zr~I/Fe]}$&$\sigma$&N& $\mathrm{[Zr~II/Fe]}$&$\sigma$ &N&$\mathrm{[Ba/Fe]}$&$\sigma$&N\\            
       \noalign{\smallskip}            
       \hline                 
        \noalign{\smallskip}            
          1   &$-$0.29 &0.10    &1       &$-$0.37 &0.08    &3       &0.18    &0.04    &2       &0.19    &0.18    &1       &$-$0.15&0.06&3\\            
        HV025 &$-$0.25 &0.10    &1       &$-$0.09 &0.09    &3       &0.11    &0.05    &2       &0.11    &0.05    &1       &$-$0.09&0.09&3\\    
          3   &$-$0.58 &0.10    &1       &$-$0.01 &0.40    &1       &$\cdots$&$\cdots$&$\cdots$&$\cdots$&$\cdots$&$\cdots$&$-$0.05&0.11&3\\
        HV007 &$-$0.20 &0.11    &1       &$-$0.20 &0.44    &2       &$\cdots$&$\cdots$&$\cdots$&$\cdots$&$\cdots$&$\cdots$&$-$0.14&0.41&3\\
        HH244 &$-$0.43 &0.11    &1       &$-$0.28 &0.14    &2       &$\cdots$&$\cdots$&$\cdots$&$\cdots$&$\cdots$&$\cdots$&$-$0.18&0.13&2\\
        HH201 &$\cdots$&$\cdots$&$\cdots$&$-$0.63 &0.20    &2       &$\cdots$&$\cdots$&$\cdots$&$\cdots$&$\cdots$&$\cdots$&$-$0.39&0.13&2\\
        HV043 &$-$0.26 &0.13    &1       &   0.49 &0.33    &1       &$\cdots$&$\cdots$&$\cdots$&$\cdots$&$\cdots$&$\cdots$&   0.34&0.21&3\\
        HV104 &$\cdots$&$\cdots$&$\cdots$&$\cdots$&$\cdots$&$\cdots$&$\cdots$&$\cdots$&$\cdots$&$\cdots$&$\cdots$&$\cdots$&$-$0.06&0.18&2\\
        HV074 &$\cdots$&$\cdots$&$\cdots$&$-$0.09 &0.32    &1       &$\cdots$&$\cdots$&$\cdots$&$\cdots$&$\cdots$&$\cdots$&$-$0.31&0.27&2\\
       \noalign{\smallskip}
       \hline
        \hline            
        \noalign{\smallskip}            
        ID of stars & $\mathrm{[La/Fe]}$ &$\sigma$ &N& $\mathrm{[Ce/Fe]}$&$\sigma$
        &N&$\mathrm{[Nd/Fe]}$&$\sigma$&N& $\mathrm{[Eu/Fe]}$*&$\sigma$&N&$\mathrm{[Dy/Fe]}$&$\sigma$ &N\\   
        \noalign{\smallskip}            
        \hline                 
         \noalign{\smallskip}            
           1   &0.34    &0.05    &2       &0.22    &0.13    &1       &0.70    &0.11    &2       &0.55    &0.11    &1       &0.56    &0.11    &1       \\            
         HV025 &0.33    &0.06    &2       &0.20    &0.16    &1       &0.37    &0.10    &2       &0.81    &0.11    &1       &0.54    &0.11    &1       \\    
           3   &$\cdots$&$\cdots$&$\cdots$&$\cdots$&$\cdots$&$\cdots$&0.40    &0.30    &1       &$\cdots$&$\cdots$&$\cdots$&$\cdots$&$\cdots$&$\cdots$\\
         HV007 &$\cdots$&$\cdots$&$\cdots$&0.65    &0.44    &1       &0.51    &0.42    &2       &$\cdots$&$\cdots$&$\cdots$&0.84    &0.42    &1       \\
         HH244 &$\cdots$&$\cdots$&$\cdots$&$\cdots$&$\cdots$&$\cdots$&0.59    &0.17    &1       &$\cdots$&$\cdots$&$\cdots$&$\cdots$&$\cdots$&$\cdots$\\
         HH201 &0.54    &0.21    &1       &$-$0.05 &0.24    &1       &$-$0.07 &0.25    &1       &$\cdots$&$\cdots$&$\cdots$&$\cdots$&$\cdots$&$\cdots$\\
         HV043 &$\cdots$&$\cdots$&$\cdots$&$\cdots$&$\cdots$&$\cdots$&0.66    &0.29    &1       &$\cdots$&$\cdots$&$\cdots$&$\cdots$&$\cdots$&$\cdots$\\
         HV104 &$\cdots$&$\cdots$&$\cdots$&$\cdots$&$\cdots$&$\cdots$&$\cdots$&$\cdots$&$\cdots$&$\cdots$&$\cdots$&$\cdots$&$\cdots$&$\cdots$&$\cdots$\\
         HV074 &$\cdots$&$\cdots$&$\cdots$&$\cdots$&$\cdots$&$\cdots$&0.71    &0.34    &1       &$\cdots$&$\cdots$&$\cdots$&$\cdots$&$\cdots$&$\cdots$\\
        \noalign{\smallskip}
        \hline
        \end{tabular}
        \tablefoot{\\ {*} The abundances of these species were derived from spectrum synthesis.}
        \end{table*}

\subsection {Comparison with the outer and inner halo GCs, extragalactic GCs
  in dSphs, and dSph field populations}

We now compare the mean abundance ratios of the nine Pal~14 stars to
those of inner and outer halo GCs, extragalactic GCs in dSphs as well as stellar abundance ratios
in dSphs in order to constrain the origin of Pal~14. We consider only GCs and dSphs that have a similar
metallicity as Pal~14.
   
\subsubsection{Palomar~14 versus M3 and M13}

\begin{figure*}[ht]
  \centering
  \includegraphics[bb=29 515 615 735,width=15.5cm,clip]{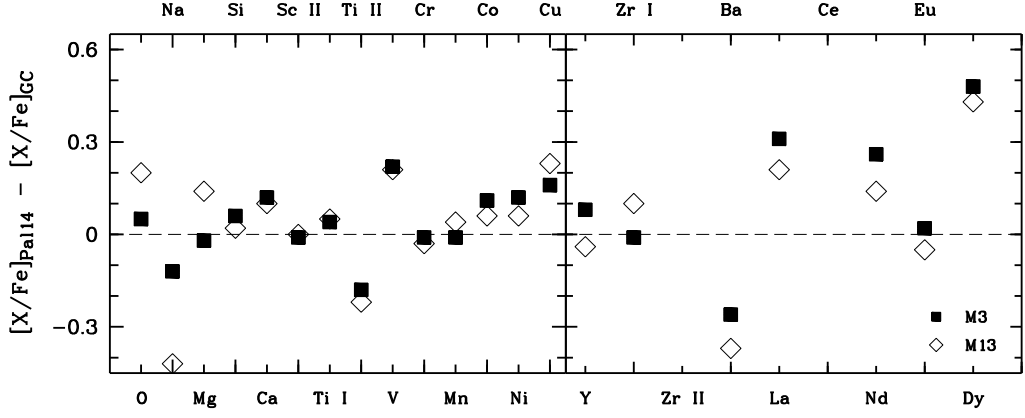}
  \caption{Difference between the mean abundance ratios of Pal~14 and
    those of the inner halo GCs M3 and M13 \citep{cohen05a}.}
  \label{inner}
\end{figure*}

\cite{cohen05a} concluded that the inner halo GCs M3 ($R_{\mathrm{GC}}\sim 9$\,kpc;
$\mathrm{[Fe/H]}=-1.39$\,dex) and M13 ($R_{\mathrm{GC}}\sim 12$\,kpc;
$\mathrm{[Fe/H]}=-1.50$\,dex), a classical second-parameter pair,
exhibit star-to-star variations among their light elements, with an
enhancement of odd atomic number elements. They found no variation in
Fe-peak elements. A $\mathrm{[Eu/Ba]}$ ratio of 0.35\,dex was
found in both clusters. This value is intermediate between the solar ratio and that
of the pure r-process, suggesting an additional r-process contribution at low metallicities.
Furthermore, \cite{cohen05a} concluded that the stars have experienced similar chemical enrichment histories compared to halo
field and other Galactic GC stars. In Figure~\ref{inner}, we show a
comparison of the abundance patterns of Pal~14 and inner halo GCs. 

Pal~14 shows almost the same abundance patterns as M3 and M13 for both the Fe-peak and
$\alpha-$elements within the abundance uncertainties. Pal~14 has
similar abundances to M3 and M13 for Y, Zr, and
Eu, while it has a lower Ba abundance and a higher abundance of La, Nd, and Dy.
The O and Mg abundance ratios of Pal~14 are about the same
as those of M3, while they are slightly higher than the abundance
ratio of M13. Therefore, the Na abundance ratios of Pal~14 are about
same as those of M3, and notably lower than in M13. The Ba and Dy
abundance ratios of Pal~14 are considerably lower and higher than
those of M3 and M13, respectively. The Cu abundance ratio is moderately
higher than that of both GCs, as is shown in Figure~\ref{inner}.

Pal~14 and the M3--M13 pair resemble each other closely in their iron and alpha elements, that is, they have experienced
similar chemical enrichment histories. It should
be pointed out that the Cu abundance ratio of Pal 14 is higher with respect to the Cu
abundance ratios of M3 and M13, while its Ba abundance ratio is lower than
the Ba abundance ratios of both M3 and M13. 
It is thought that explosive nucleosynthesis in supernovae
(type~Ia and type~II) contributes to the production of Cu, which is a
transition element between the Fe-peak and light s-process elements
\citep{snedenetal91,misheninaetal02}. Barium is
mostly produced by s-process nucleosynthesis, occuring in AGB stars during the thermal pulse
instability phase \citep{burrisetal00}. Therefore, the contribution
coming from SN~II might be higher, even if both SN~Ia and SN~II
contributed to the chemical enrichment of Pal~14 with respect to M3
and M13, because of higher Cu and lower Ba abundance ratios of Pal~14.

\subsubsection{Palomar~14 versus other outer halo globular clusters}

We now compare Pal~14 with Pal~3 \citep[$R_{\mathrm{GC}}\sim 96$\,kpc,
$\mathrm{[Fe/H]}=-1.58$\,dex;][]{kochetal09}, Pal~4 \citep[$R_{\mathrm{GC}}\sim 110$\,kpc,
$\mathrm{[Fe/H]}=-1.41$\,dex;][]{koch10}, NGC~7492 \citep[$R_{\mathrm{GC}}\sim 25$\,kpc,
$\mathrm{[Fe/H]}=-1.87$\,dex;][]{cohen05b}, and NGC~7006 \citep[($R_{\mathrm{GC}}\sim 40$\,kpc,
$\mathrm{[Fe/H]}=-1.55$\,dex;][]{kraftetal98}.

\cite{kochetal09} report on an abundance analysis of four giant stars
of Pal~3. They found that the abundance pattern of Pal~3 agrees better
with Galactic halo field stars and Galactic globular clusters than with dwarf spherodial(dSph) galaxies. No conclusive evidence of
star-to-star variations of any chemical element in this distant GC has
been found. This result is complemented with an abundance analysis of Pal~4 \citep{koch10} based on co-added
spectra.

An abundance analysis of again four giant stars in NGC~7492 was
carried out by \citet{cohen05b}. For most elements, no star-to-star
variations were found, but this GC shows a Na--O anticorrelation and
an abundance pattern similar to the inner halo GCs. Moreover, the
neutron-capture abundance pattern of NGC~7492 indicated that both r and
s-process enrichment proceeded in a similar fashion as in M3 and M13. 
In NGC~7006, six stars were analyzed by \cite{kraftetal98} and the iron peak and $\alpha-$element ratios are
similar to those found in the other halo GCs. This cluster exhibits star-to-star
abundance variations in O, Na, and Al.

\begin{figure*}[ht]
  \centering
  \includegraphics[bb=29 515 615 735,width=15.5cm,clip]{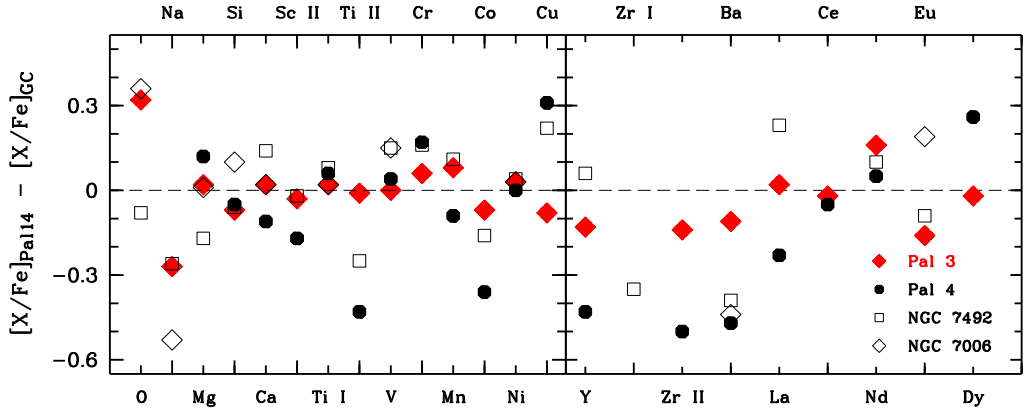}
  \caption{Difference between the mean abundance ratios of Pal~14 and
    the outer halo GCs Pal~3, Pal~4, NGC~7006, and NGC~7492. For references, see text.}
  \label{outer} 
\end{figure*}

In Figure~\ref{outer} we show the differences between Pal~14 and the
aforementioned GCs. Pal~14 is enhanced in O relative the other
clusters except for NGC~7492, nonetheless it is depleted in the Na
abundance ratio. In the iron peak and $\alpha-$elements,
Pal~14 agrees best with Pal~3, although it also agrees with the other
clusters.

One of the remarkable points is the similarity between Pal~14
and Pal~3 in their neutron-capture elements. This would be an
indicator as to their similar origin. Generally speaking, Pal~14 appears to be
more compatible with the outer halo, in particular with the most distant outer halo GCs as compared to the inner
halo GCs. That is, their chemical enrichment histories appear to be
fairly similar. However, abundance analyses based on higher quality
data are needed to confirm this.

\subsubsection{Palomar 14 versus extragalactic GCs in dSphs} 
\label{sec:extragalacticgcs}

\begin{figure*}[ht]
  \centering
  \includegraphics[bb=29 515 615 735,width=15.5cm,clip]{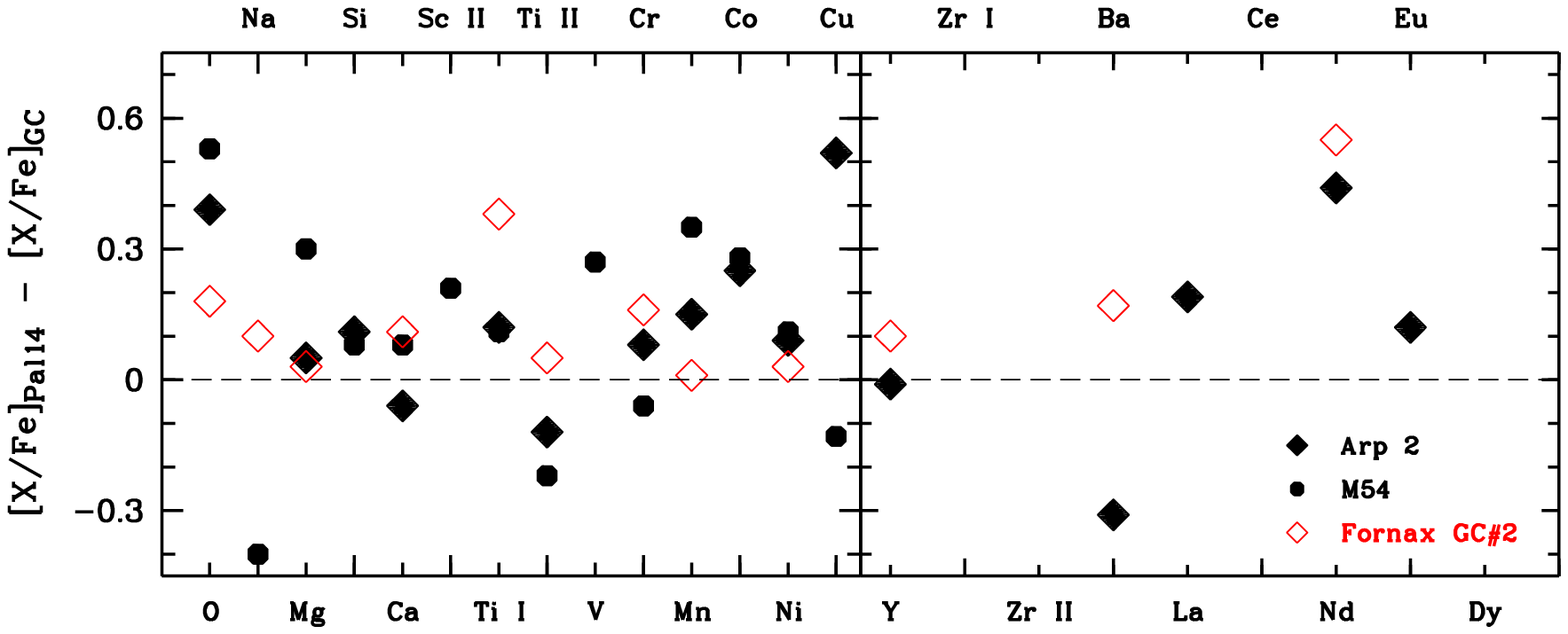}
  \caption{Difference of the abundance ratios in between Pal 14 and
    Arp~2 \citep{mottinietal08}, Fornax~GC$\#$2 \citep{letarteetal06}, and M54 \citep{carrettaetal10}.}
  \label{extragcs} 
\end{figure*}

Following the discovery of the accreted dSph Sagittarius (Sgr) \citep{ibataetal94}, some Galactic GCs
(e.g, Pal~12, Ter~7, Arp~2) were associated with it
\citep{cohen04,sbordoneetal07,mottinietal08}. Apart from the Sgr dSph,
  GCs were detected only in one additional Galactic dSph, Fornax. Fornax
  contains five GCs. To investigate the possible
extragalactic origin of Pal~14, we compare Pal~14 with Arp~2
\citep[$\mathrm{[Fe/H]}=-1.77$\,dex;][]{mottinietal08}, M54
\citep[$\mathrm{[Fe/H]}=-1.56$\,dex;][]{carrettaetal10}, and Fornax~GC$\#$2
\citep[$\mathrm{[Fe/H]}=-2.10$\,dex;][]{letarteetal06} of these GCs,
since the chosen comparison clusters have a similar metallicity as Pal~14.
\citet{mottinietal08} performed a chemical abundance analysis of Arp~2 (two
stars). They found that the iron peak and $\alpha-$element ratios of Arp~2 are
not too different from the Galactic GCs that have similar
metallicities. \citet{carrettaetal10} derived the abundances of
  the light elements, $\alpha-$elements, and iron peak elements of 76 red
  giant stars in M54 which is located in the Sgr dSph. In the study of \citet{letarteetal06} the abundance
analysis of three Fornax GCs is reported. These authors also concluded that the abundance pattern
of GCs in Fornax and Milky Way are quite similar and that both Milky Way and
Fornax globular clusters shared the same initial conditions when they
formed. In Figure~\ref{extragcs}, we show the differences between Pal~14 and
the aforementioned Sgr and Fornax GCs. Pal~14 exhibits similar abundance
patterns as Arp~2, M54, and Fornax~GC$\#$~2 in the iron peak and
$\alpha-$elements. It is worthy of note that the Y abundance of
Pal~14 is same as in Fornax~GC$\#$2 and the light element abundances of
  Pal~14 are different from those of M54.

\subsubsection{Palomar 14 in comparison to dSph field populations and
    ultra-faint dSphs}

For a comparison between Pal~14 and dSphs, we choose three dSphs stars
with metallicities similiar to that of Pal~14, and
use the mean abundances of the stars of each dSph. The dSphs are Ursa
Minor with a Galactocentric distance of
$R_{\mathrm{GC}}\sim 70$\,kpc and $\mathrm{[Fe/H]}=-1.54$
\citep{cohen10}, Draco with $R_{\mathrm{GC}}\sim80$\,kpc and
$\mathrm{[Fe/H]}=-1.60$ \citep{cohen09}, and Sculptor
with $R_{\mathrm{GC}}\sim86$\,kpc, $\mathrm{[Fe/H]}=-1.52$
\citep{shetroneetal03}. All three dSphs are dominated by old populations
that are of comparable age as the ages of the oldest Galactic GCs 
\citep{grebel04}.  \citet{cohen10} suggested that the lower-metallicity
stars in both UMi and Dra are similar to the outer Galactic
halo stars in their $\alpha$ and iron peak elements, while their abundance ratios
are lower than those of halo stars at higher metallicities. As for the neutron-capture elements, they exhibit an r-process
distribution at $\mathrm{[Fe/H]}\le-2$, switching to an s-process
distribution at the highest metallicities. This suggests that both the inner and outer Galactic halo may have
formed via accretion of dSphs, but the inner halo accretion event
occured earlier than that of the outer halo. Furthermore, the
dissolution of GCs might have contributed to the inner halo (see, e.g.,
\citealt{jordi10,martell10}).
\citet{shetroneetal03} suggested that the dSphs show Galactic halo
like abundances in the iron peak elements. Also, the $\alpha-$element abundances of the dSphs
may vary from galaxy to galaxy, but are lower than in Galactic
halo stars with similar metallicity. That is, the bulk of the Galactic
halo cannot have formed via accretion of systems like the present-day
dSphs, even though it is likely that there were many early contributions. 

\begin{figure*}[ht]
  \centering
  \includegraphics[bb=29 515 615 735,width=15.5cm,clip]{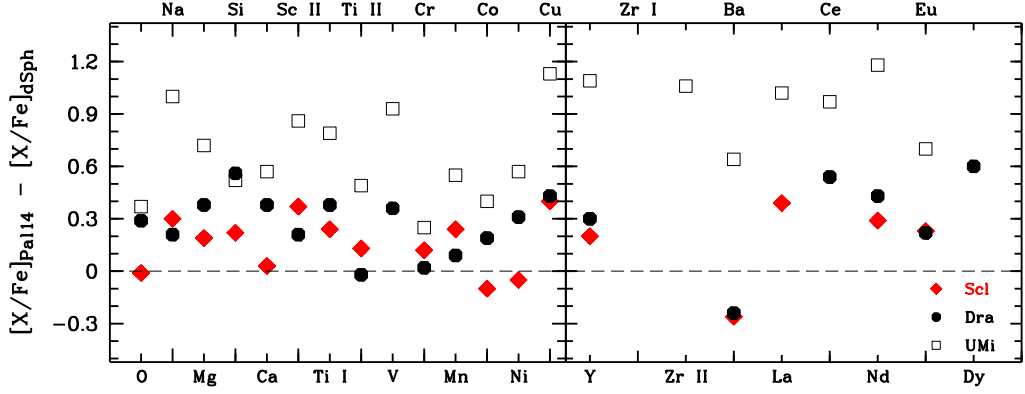}
  \caption{Difference of the abundance ratios in between Pal 14 and
    the individual stars with similar metallicity in the UMi \citep{cohen10},
    Dra \citep{cohen09}, and
    Scl \citep{shetroneetal03} dSphs.}
  \label{dSph} 
\end{figure*}

As shown in Figure~\ref{dSph}, Pal~14 exhibits higher abundance ratios with respect
to those of the UMi dSph stars, but it does not differ significantly from Scl
and Dra stars with similar metallicity.

Furthermore, and not surprisingly, Pal 14 differs greatly from typical dSphs in
velocity dispersion ($\sigma = 0.38$\,$\mathrm{km~s^{-1}}$; \citealt{jordietal09}) as
well as mass-to-light ratio ($\sim2$ in solar units;
\citealt{jordietal09}), since dSphs typically have velocity
dispersions of $\sim 8$--$15$\,$\mathrm{km~s^{-1}}$ and
mass-to-light ratios of $\ge 30$, which is usually attributed to the presence
of dark matter in dSphs \citep{gilmoreetal07}.

Another group in the Galactic halo are ultra-faint dSphs (UF-dSphs) which are
fainter and more metal-poor than classical dSphs
\citep{simon07,kirbyetal08,gehaetal09,frebeletal10,norrisetal10,adenetal11}. Although
UF-dSphs have similar or lower total luminosities as globular clusters,
UF-dSphs seem to be dark matter dominated dSphs
\citep{martinetal07,simon07,gehaetal09,adenetal09}. Furthermore, those objects
show internal metallicity spreads up to 0.5\,dex contrast to globular clusters
\citep{simon07,frebeletal10}. Given these properties, Pal~14 with essentially
a single metallicity and a low mass-to-light ratio, like Pal~3 and Pal~4, does
not resemble the UF-dSphs. Also, Pal~14, with a half-light radius of
46.1$\pm$2.9\,pc \citep{sollimaetal10}, is a GC, not an extended diffuse
object like an UF-dSph \citep[see, e.g., Fig.1][]{misgeld11}.

\subsection{Neutron-capture element abundance patterns} 

\begin{figure}[ht]
  \centering
  \includegraphics[bb=37 191 553 751,width=9cm,clip]{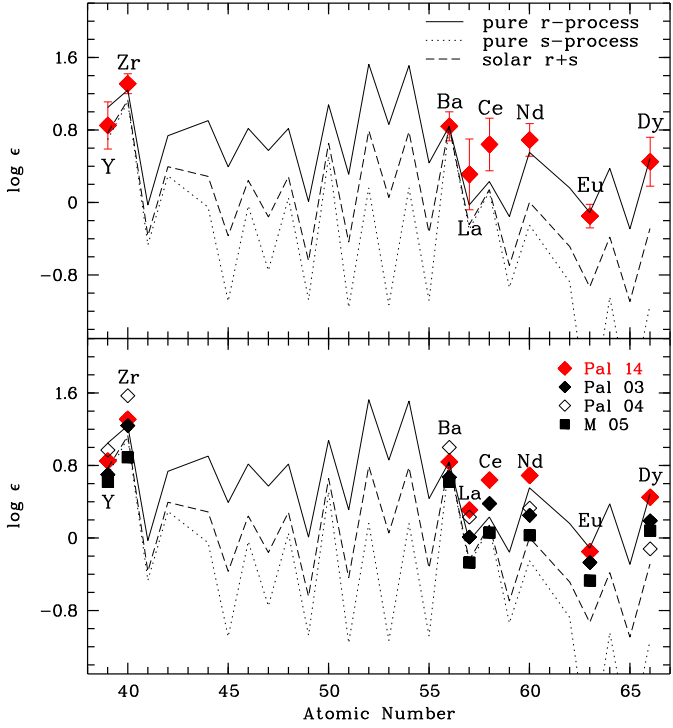}
  \caption{The mean abundance ratios of neutron-capture elements in
    Pal~14. We compared to the solar scaled r, s and r+s-process
    abundance ratios \citep{burrisetal00}, normalised to Ba. The error bars
    indicate the $\sigma-$spread for each element. M5 data are taken from \citet{laietal11}.}
  \label{neutcapture}
\end{figure}

In Figure~\ref{neutcapture}, we show the mean abundances for
individual neutron-capture elements of Pal~14 compared to the Solar scaled abundances of
\citet{burrisetal00}, scaled to match the Ba abundance of Pal~14. 

The mean neutron-capture element abundances of
Pal~14 agree well with the scaled solar r-process abundance
pattern. To understand whether this abundance pattern shows a
star-to-star variation, we checked whether star~1 and HV025, which have all neutron-capture
elements detected in our spectra, are enhanced with r-process nucleosynthesis. The result was still
the same, i.e., the heavy element production in the putative proto-cluster was governed
by the r-process associated with contributions of massive SNe~II rather
than the s-process related to the contribution of AGB stars \citep{truran88,snedenetal97,otsukuetal06}.

We note that Pal~14 is the fourth cluster in which an r-process
abundance signature was found, the other three being M15, Pal~3, and
M5. \citet{snedenetal97,snedenetal00} found that M15 exhibits
star-to-star heavy element abundance variation and the heavy element abundance
is consistent with pure r-process nucleosynthesis. Later, these results were confirmed by
\citet{otsukuetal06}. They also suggested that only a single r-process
contribution is not enough to explain the observed light neutron-capture
element abundances in M15. In connection with their study of Pal~3,
\citet{kochetal09} pointed out that the r-process might have occured
without any need to invoke an enrichment by very massive stars, as was found in
ultra-faint dSph galaxies. \citet{laietal11} found that the neutron-capture
element abundances of M5 show predominantly an r-process signature, but the
cluster has a small uniform addition of s-process material. They also
emphasized that the neutron-capture signature is the same for all stars in
their sample and interpreted the result as indicating that low-mass AGB stars
contributed heavy elements to the primordial cluster environment. Then,
\citet{roederer11} found a correlation between the the $\mathrm{[La/Eu]}$ and
$\mathrm{[Eu/Fe]}$ ratios in M5 re-examining heavy element abundances of the
cluster and reported that M5 shows an r-process dispersion.

\begin{figure}[ht]
  \centering
  \includegraphics[bb=37 191 553 751,width=9cm,clip]{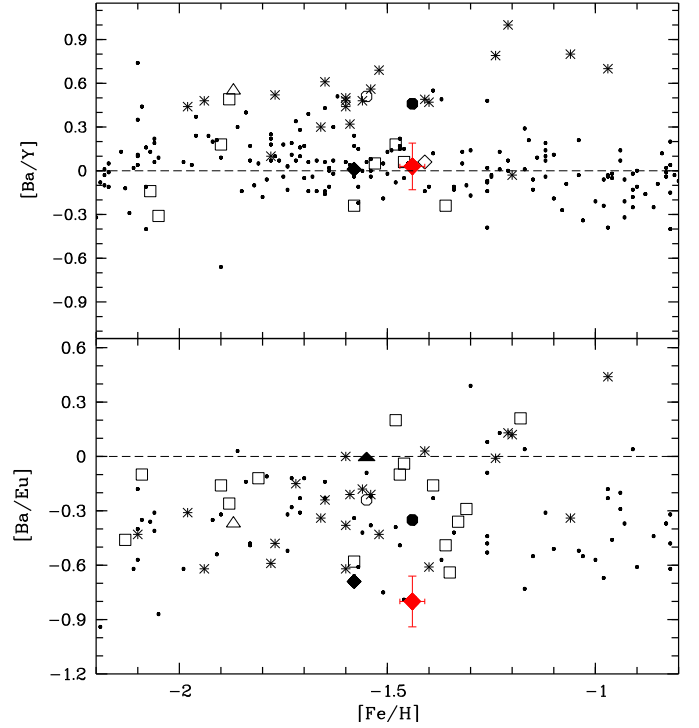}
  \caption{[Ba/Y] and [Ba/Eu] abundance ratios as a function of
    [Fe/H]. The symbols and references are the same as in Figure~\ref{alpha}.}
   \label{bayeu} 
   \end{figure}

\citet{vennetal04} found that the $\mathrm{[Ba/Y]}$ ratio in metal-poor dSph stars is
higher than in Galactic stars of similar metallicity because of the contributions
from metal-poor AGB stars \citep{travaglioetal04}. Therefore, this ratio is a useful
tool to compare the Galactic stars and GCs with the dSph stars. As shown in
Figure~\ref{bayeu}, the $\mathrm{[Ba/Y]}$ abundance ratio of Pal~14 is
compatible with the other outer and the inner halo GCs, and is slightly lower compared
to dSph stars at similar metallicity.  

The $\mathrm{[Ba/Eu]}$ abundance ratio is a good indicator of which neutron-capture process is
dominant in the production of heavy elements. Pal~14 exhibits a slightly lower 
$\mathrm{[Ba/Eu]}$ ratio compared to halo stars, GCs, and dSphs at similar
metallicity, as seen in Figure~\ref{bayeu}. This ratio is
another evidence of r-process nucleosynthesis in Pal~14. 

\section{Conclusions}

We derived the chemical abundances of nine stars belonging to Pal~14,
using high-resolution spectra with moderate $S/N$ ratio, and compared
the resulting abundances to inner and outer halo GCs and extragalactic GCs
in dSphs as well as dSphs and halo
field stars. Pal~14 exhibits an abundance pattern that is almost identical to that of
Pal~3, which is one of the most distant outer halo GCs. Both GCs are
also similar in metallicity, age, distance from the Galactic center,
and diffuseness. Hence, they seem to have experienced similar
formation and chemical enrichment histories. The abundance patterns of Pal~14 and Pal~3 are also similar
to those of the other inner and outer halo GCs, extragalactic GCs in
  dSphs, and dSph field populations as well as halo field stars
with similar metallicity, except for the neutron-capture elements. 

Our limited data quality does not allow us to determine whether the
Na-Ni relation, which is expected when SNe~II make a dominant contribution to the
chemical enrichment \citep{vennetal04,letarteetal10}, is compatible with that is seen in halo field stars.

Regarding neutron-capture elements, Pal~14 is consistent with an r-process
nucleosynthesis scenario. The $\mathrm{[Ba/Eu]}$ abundance ratio of Pal~14 is moderately low compared to 
halo field stars, GCs, and dSphs, while it is consistent with Pal~3. The average
$\mathrm{[Ba/Y]}$ ratio of nine Pal~14 giants is slightly increased
compared to halo field stars.

Our data do not permit us to investigate the Na and O abundance variations
among Pal~14 stars. Nonetheless, we discuss the possibility of light element
abundance variations. In the case of the existence of light element abundance
variations, our results indicate that in its early formation history, Pal~14 may have
contained its primordial massive stars in the progenitor cloud's central region, and these stars exploded
as SNe~II, thus the interstellar medium (ISM) of Pal~14 could have
been enriched by iron, $\alpha$, and neutron-capture elements, particularly
r-process elements \citep{truran88,parmentier04}, in contrast to the
self-enrichment scenario. The self-enrichment scenario suggests that GCs were uniformly
enriched by r and s-process nucleosynthesis, and that most
of them show the same abundance pattern as halo field stars
\citep{jamesetal04}. This classical self-enrichment may explain why Pal~14
bears the traces of r-process nucleosynthesis. Its first generation
stars \citep[the primordial component of][]{carrettaetal09a,carrettaetal09b} -- massive or intermediate- and
low-mass AGB or massive binary stars -- might form out of polluted gas
\citep{deminketal10,dercoleetal08,decressinetal07,prantzos06}. These stars
produced light elements during their evolution and the ISM of the cluster became more
enriched with the products of H-burning. The next generation could have been formed in
this environment. Thus, these different generation stars are responsible for the observed abundance
variations. We suggest that: (1) the $\mathrm{[Mg/Ca]}$ abundance ratios of Pal~14 giants are enriched by
SNe~II with progenitor masses in the range $15$--$23$\,M$_{\odot}$
\citep{heger10}, (2) Pal~14 is enhanced by the r-process elements in heavy elements. 

We report that Pal~14 has no suprising abundance patterns when compared with
those of the outer Galactic halo GCs and bears a striking resemblance to
Pal~3. Moreover, while Pal~14 differs from the abundance trends in observed in
dSph field stars, it closely resembles the trends found in the few
dSph GCs studied in detail so far. We thus conclude that both
the other outer halo GCs as well as Pal 14 may well have had their
origin in since accreted dSphs or their progenitors. Whether or not GC accretion played a role,
it seems that the formation conditions of outer halo GCs and GCs in dSphs were
similar.

Our findings regarding the light element abundance variations and neutron-capture elements pattern in Pal~14
need to be confirmed with high-quality data. Additional data would help to improve our
understanding not only of the formation and evolution of GCs but also of the outer Galactic halo.
        
\begin{acknowledgements}
      
The authors thank an anonymous referee for the thoughtful comments. We thank
Katrin Jordi for providing us with reduced and co-added
FLAMES/UVES spectra of the nine Pal~14 giants. We are grateful to Hans
Ludwig, Elisebetta Caffau, Andreas Koch, Sarah Martell, and Luca Sbordone for their
help and advice. Andreas Koch and Robert L.\ Kurucz kindly provided us
with line lists, which is gratefully acknowledged. \c{S}\c{C} thanks the Higher
Educational Council of Turkey (Y\"{O}K) for the financial support during this
project. This work was supported by Sonderforschungsbereich SFB 881 ''The Milky
Way System'' (subproject A5) of the German Research Foundation (DFG).

\end{acknowledgements}

\bibliographystyle{aa}
\bibliography{palomar14}

\begin{appendix}
\section{Line list, equivalent width measurements, abundances derived for each line.}

\onllongtab{1}{
\tiny
\begin{longtable}{lccrrrrrrrrrrr}
\caption{\label{Tab:Linelist} Line data, equivalent widths, and abundances
  derived for each line for the sample of Pal~14 giants.}\\
\hline
\hline
Species& $\lambda$ & $\chi$ & $\log gf$ & EW01 & $A$ &EWHV025&$A$ &EW03&$A$ & EWHV007&$A$&EWHH244&$A$ \\
&[{\AA}]&[eV]&&[pm]&&[pm]&&[pm]&&[pm]&&[pm]&\\
\hline
\endfirsthead
\caption{Continued.} \\
\hline
Species& $\lambda$ & $\chi$ & $\log gf$ & EW01 & $A$ &EWHV025&$A$&EW03&$A$ & EWHV007&$A$&EWHH244&$A$ \\
&[{\AA}]&[eV]&&[pm]&&[pm]&&[pm]&&[pm]&&[pm]&\\
\hline
\endhead
\hline
\endfoot
\hline
\endlastfoot
O~I   & 6300.33 & 0.00 & $-$9.82 &syn     &7.83    &syn     &7.85    &syn     &8.24    &syn     &8.08    &syn     &8.15\\
O~I   & 6363.76 & 0.02 & $-$10.30&syn     &7.86    &syn     &7.99    &$\cdots$&$\cdots$&$\cdots$&$\cdots$&$\cdots$&$\cdots$\\
Na~I  & 5682.64 & 2.10 & $-$0.70 &5.57    &4.64    &6.5     &4.79    &$\cdots$&$\cdots$&$\cdots$&$\cdots$&$\cdots$&$\cdots$\\
Na~I  & 5688.21 & 2.10 & $-$0.42 &7.22    &4.58    &7.57    &4.64    &3.90    &4.47    &7.87    &4.94    &5.58    &4.57    \\
Mg~I  & 5528.42 & 4.35 & $-$0.36 &syn     &6.71    &syn     &6.48    &16.53   &6.13    &syn     &6.64    &syn     &6.64    \\
Mg~I  & 5711.11 & 4.33 & $-$1.73 &12.93   &6.71    &9.5     &6.22    &8.84    &6.46    &8.09    &6.29    &9.15    &6.34    \\
Si~I  & 5690.47 & 4.93 & $-$1.77 &2.54    &6.21    &3.51    &6.40    &$\cdots$&$\cdots$&$\cdots$&$\cdots$&$\cdots$&$\cdots$\\
Si~I  & 5948.55 & 5.08 & $-$0.78 &6.73    &6.15    &7.2     &6.17    &9.37    &6.52    &8.93    &6.41    &$\cdots$&$\cdots$\\
Si~I  & 6155.16 & 5.61 & $-$0.75 &3.16    &6.22    &3.88    &6.33    &$\cdots$&$\cdots$&4.59    &6.44    &$\cdots$&$\cdots$\\
Ca~I  & 5261.71 & 2.52 & $-$0.58 &11.21   &5.02    &11.39   &5.02    &12.93   &5.74    &$\cdots$&$\cdots$&$\cdots$&$\cdots$\\
Ca~I  & 5581.97 & 2.53 & $-$0.56 &10.55   &4.83    &11.68   &5.00    &11.03   &5.35    &10.21   &5.11    &$\cdots$&$\cdots$\\
Ca~I  & 5601.28 & 2.53 & $-$0.52 &12.41   &5.11    &12.74   &5.13    &9.58    &5.10    &13.38   &5.59    &13.23   &5.49    \\
Ca~I  & 5857.46 & 2.93 &    0.24 &13.44   &5.01    &15.48   &5.27    &13.71   &5.44    &13.25   &5.26    &10.67   &4.80    \\
Ca~I  & 6122.24 & 1.89 & $-$0.32 &$\cdots$&$\cdots$&$\cdots$&$\cdots$&$\cdots$&$\cdots$&$\cdots$&$\cdots$&$\cdots$&$\cdots$\\
Ca~I  & 6166.45 & 2.52 & $-$1.14 &9.07    &5.12    &9.58    &5.20    &7.48    &5.37    &5.52    &5.02    &6.28    &5.02    \\
Ca~I  & 6169.05 & 2.52 & $-$0.80 &10.25   &4.95    &11.44   &5.12    &8.70    &5.20    &8.35    &5.05    &9.01    &5.04    \\
Ca~I  & 6169.56 & 2.52 & $-$0.48 &12.52   &4.99    &13.12   &5.05    &8.37    &4.84    &12.21   &5.28    &8.32    &4.63    \\
Ca~I  & 6439.09 & 2.53 &    0.39 &18.39   &4.95    &18.23   &4.87    &$\cdots$&$\cdots$&$\cdots$&$\cdots$&$\cdots$&$\cdots$\\
Ca~I  & 6455.59 & 2.89 & $-$1.29 &7.88    &5.08    &8.01    &5.12    &8.49    &5.64    &8.02    &5.48    &$\cdots$&$\cdots$\\
Ca~I  & 6471.67 & 2.52 & $-$0.69 &12.98   &5.22    &14.21   &5.37    &11.40   &5.47    &9.76    &5.30    &$\cdots$&$\cdots$\\
Ca~I  & 6499.63 & 2.52 & $-$0.82 &8.81    &4.73    &7.91    &4.64    &10.87   &5.51    &$\cdots$&$\cdots$&$\cdots$&$\cdots$\\
Ca~I  & 6572.78 & 0.00 & $-$4.24 &12.39   &4.85    &11.86   &4.87    &9.65    &5.54    &$\cdots$&$\cdots$&$\cdots$&$\cdots$\\
Ca~I  & 6717.69 & 2.71 & $-$0.52 &13.07   &5.30    &14.1    &5.43    &13.68   &5.86    &12.87   &5.70    &13.66   &5.66    \\
Sc~II & 5239.84 & 1.46 & $-$0.77 &9.16    &1.78    &8.24    &1.60    &11.51   &2.37    &8.45    &1.79    &9.85    &1.96    \\
Sc~II & 5526.83 & 1.77 &    0.02 &11.78   &1.82    &10.39   &1.53    &12.55   &2.09    &$\cdots$&$\cdots$&$\cdots$&$\cdots$\\
Sc~II & 5669.03 & 1.50 & $-$1.20 &6.42    &1.82    &7.83    &2.00    &6.86    &2.13    &8.89    &2.31    &$\cdots$&$\cdots$\\
Sc~II & 5684.22 & 1.51 & $-$1.07 &8.62    &2.03    &10.53   &2.27    &9.83    &2.43    &9.35    &2.25    &6.77    &1.85    \\
Sc~II & 6245.62 & 1.51 & $-$1.03 &9.91    &2.12    &9.16    &1.98    &8.34    &2.13    &8.21    &2.01    &9.51    &2.15    \\
Sc~II & 6604.61 & 1.36 & $-$1.31 &9.62    &2.12    &7.48    &1.81    &7.68    &2.11    &$\cdots$&$\cdots$&10.31   &2.31    \\
Ti~I  & 4997.10 & 0.00 & $-$2.12 &13.78   &3.63    &13.13   &3.55    &12.21   &4.50    &$\cdots$&$\cdots$&$\cdots$&$\cdots$\\
Ti~I  & 4999.51 & 0.83 &    0.25 &$\cdots$&$\cdots$&$\cdots$&$\cdots$&15.34   &3.77    &$\cdots$&$\cdots$&$\cdots$&$\cdots$\\
Ti~I  & 5009.66 & 0.02 & $-$2.26 &14.83   &4.00    &15.07   &4.03    &$\cdots$&$\cdots$&$\cdots$&$\cdots$&$\cdots$&$\cdots$\\
Ti~I  & 5022.89 & 0.83 & $-$0.43 &16.81   &3.78    &$\cdots$&$\cdots$&$\cdots$&$\cdots$&$\cdots$&$\cdots$&$\cdots$&$\cdots$\\
Ti~I  & 5045.43 & 0.85 & $-$2.00 &6.28    &3.68    &7.24    &3.90    &$\cdots$&$\cdots$&$\cdots$&$\cdots$&$\cdots$&$\cdots$\\
Ti~I  & 5062.13 & 2.16 & $-$0.46 &5.25    &3.89    &4.55    &3.86    &$\cdots$&$\cdots$&$\cdots$&$\cdots$&$\cdots$&$\cdots$\\
Ti~I  & 5147.49 & 0.00 & $-$1.88 &14.54   &3.46    &15.28   &3.59    &8.05    &3.59    &$\cdots$&$\cdots$&11.72   &3.68    \\
Ti~I  & 5152.18 & 0.02 & $-$1.91 &14.94   &3.60    &$\cdots$&$\cdots$&13.89   &4.57    &11.6    &3.97    &$\cdots$&$\cdots$\\
Ti~I  & 5210.39 & 0.05 & $-$0.88 &$\cdots$&$\cdots$&$\cdots$&$\cdots$&16.46   &4.00    &$\cdots$&$\cdots$&18.01   &3.72    \\
Ti~I  & 5219.71 & 0.02 & $-$1.98 &16.23   &3.88    &15.17   &3.68    &8.43    &3.77    &11.26   &3.97    &8.78    &3.39    \\
Ti~I  & 5401.35 & 0.82 & $-$2.89 &2.76    &3.95    &$\cdots$&$\cdots$&$\cdots$&$\cdots$&$\cdots$&$\cdots$&$\cdots$&$\cdots$\\
Ti~I  & 5644.13 & 2.27 &    0.21 &7.32    &3.59    &6.84    &3.58    &4.05    &3.77    &$\cdots$&$\cdots$&$\cdots$&$\cdots$\\
Ti~I  & 5648.54 & 2.50 & $-$0.26 &$\cdots$&$\cdots$&3.48    &3.90    &$\cdots$&$\cdots$&$\cdots$&$\cdots$&$\cdots$&$\cdots$\\
Ti~I  & 5866.46 & 1.07 & $-$0.84 &11.70   &3.44    &13.94   &3.79    &8.67    &3.93    &$\cdots$&$\cdots$&$\cdots$&$\cdots$\\
Ti~I  & 5918.55 & 1.07 & $-$1.46 &7.34    &3.49    &7.01    &3.54    &6.20    &4.22    &$\cdots$&$\cdots$&$\cdots$&$\cdots$\\
Ti~I  & 5922.11 & 1.05 & $-$1.47 &9.69    &3.75    &9.16    &3.76    &4.66    &4.00    &5.22    &3.94    &3.67    &3.52    \\
Ti~I  & 5941.76 & 1.05 & $-$1.51 &8.48    &3.65    &7.76    &3.65    &3.24    &3.83    &$\cdots$&$\cdots$&$\cdots$&$\cdots$\\
Ti~I  & 5965.82 & 1.88 & $-$0.41 &8.11    &3.72    &7.74    &3.73    &2.74    &3.66    &8.37    &4.35    &3.77    &3.57    \\
Ti~I  & 5978.54 & 1.88 & $-$0.50 &7.13    &3.67    &7.64    &3.80    &3.49    &3.88    &$\cdots$&$\cdots$&$\cdots$&$\cdots$\\
Ti~I  & 6064.62 & 1.05 & $-$1.94 &4.33    &3.54    &5.47    &3.79    &2.92    &4.19    &$\cdots$&$\cdots$&$\cdots$&$\cdots$\\
Ti~I  & 6126.22 & 1.07 & $-$1.42 &10.85   &3.86    &10.27   &3.86    &3.95    &3.86    &5.35    &3.92    &$\cdots$&$\cdots$\\
Ti~I  & 6258.12 & 1.44 & $-$0.36 &13.23   &3.66    &12.66   &3.62    &12.70   &4.46    &10.62   &3.99    &7.02    &3.36    \\
Ti~I  & 6303.77 & 1.44 & $-$1.57 &5.27    &3.84    &5.11    &3.92    &$\cdots$&$\cdots$&$\cdots$&$\cdots$&$\cdots$&$\cdots$\\
Ti~I  & 6554.25 & 1.44 & $-$1.22 &6.65    &3.65    &6.46    &3.71    &2.55    &3.86    &$\cdots$&$\cdots$&$\cdots$&$\cdots$\\
Ti~I  & 6556.07 & 1.46 & $-$1.07 &9.08    &3.83    &7.42    &3.71    &5.07    &4.14    &$\cdots$&$\cdots$&4.84    &3.67    \\
Ti~I  & 6599.10 & 0.90 & $-$2.09 &6.96    &3.75    &8.02    &3.97    &$\cdots$&$\cdots$&$\cdots$&$\cdots$&$\cdots$&$\cdots$\\
Ti~I  & 6743.15 & 0.90 & $-$1.63 &8.72    &3.48    &8.78    &3.58    &3.05    &3.67    &$\cdots$&$\cdots$&7.40    &3.91    \\
Ti~II & 5005.16 & 1.57 & $-$2.72 &5.35    &3.70    &6.1     &3.81    &$\cdots$&$\cdots$&$\cdots$&$\cdots$&$\cdots$&$\cdots$\\
Ti~II & 5013.70 & 1.58 & $-$2.19 &8.86    &3.76    &9.39    &3.79    &$\cdots$&$\cdots$&$\cdots$&$\cdots$&11.29   &3.94    \\
Ti~II & 5185.91 & 1.89 & $-$1.35 &12.99   &4.04    &15.75   &4.38    &12.19   &4.01    &14.67   &4.30    &14.89   &4.35    \\
Ti~II & 5226.54 & 1.57 & $-$1.26 &15.50   &3.93    &14.89   &3.71    &16.29   &4.21    &17.36   &4.24    &14.39   &3.79    \\
Ti~II & 5336.81 & 1.58 & $-$1.70 &$\cdots$&$\cdots$&$\cdots$&$\cdots$&$\cdots$&$\cdots$&$\cdots$&$\cdots$&14.09   &4.14    \\
Ti~II & 5396.27 & 1.58 & $-$2.51 &5.56    &3.52    &5.93    &3.57    &5.56    &3.76    &9.55    &4.23    &$\cdots$&$\cdots$\\
Ti~II & 5418.79 & 1.58 & $-$2.00 &9.29    &3.60    &11.53   &3.89    &12.88   &4.36    &8.34    &3.55    &9.02    &3.61    \\
Ti~II & 6559.55 & 2.05 & $-$2.02 &4.96    &3.52    &5.92    &3.65    &5.09    &3.72    &4.48    &3.56    &$\cdots$&$\cdots$\\
V~I   & 5627.64 & 1.08 & $-$0.36 &9.82    &2.77    &7.61    &2.56    &5.47    &3.05    &4.97    &2.85    &5.99    &2.79    \\
V~I   & 5670.85 & 1.08 & $-$0.42 &5.31    &2.23    &7.94    &2.65    &$\cdots$&$\cdots$&$\cdots$&$\cdots$&$\cdots$&$\cdots$\\
V~I   & 5703.58 & 1.05 & $-$0.21 &10.00   &2.58    &10.17   &2.67    &$\cdots$&$\cdots$&$\cdots$&$\cdots$&$\cdots$&$\cdots$\\
V~I   & 6039.73 & 1.06 & $-$0.65 &5.35    &2.40    &5.12    &2.47    &2.21    &2.75    &4.36    &3.00    &4.29    &2.80    \\
V~I   & 6081.47 & 1.05 & $-$0.58 &7.23    &2.55    &4.52    &2.30    &5.54    &3.21    &4.17    &2.88    &6.12    &2.95    \\
V~I   & 6090.21 & 1.08 & $-$0.06 &9.33    &2.33    &10.12   &2.50    &$\cdots$&$\cdots$&$\cdots$&$\cdots$&$\cdots$&$\cdots$\\
V~I   & 6111.66 & 1.04 & $-$0.72 &5.19    &2.41    &6.74    &2.70    &$\cdots$&$\cdots$&$\cdots$&$\cdots$&$\cdots$&$\cdots$\\
V~I   & 6135.38 & 1.05 & $-$0.75 &6.08    &2.57    &4.38    &2.44    &$\cdots$&$\cdots$&$\cdots$&$\cdots$&$\cdots$&$\cdots$\\
V~I   & 6216.37 & 4.73 & $-$1.42 &11.71   &2.74    &12.72   &2.93    &$\cdots$&$\cdots$&$\cdots$&$\cdots$&$\cdots$&$\cdots$\\
V~I   & 6224.51 & 0.29 & $-$2.01 &6.08    &2.68    &4.7     &2.63    &$\cdots$&$\cdots$&$\cdots$&$\cdots$&$\cdots$&$\cdots$\\
V~I   & 6243.11 & 0.30 & $-$0.98 &13.85   &2.60    &14.53   &2.74    &5.89    &2.68    &5.29    &2.45    &6.53    &2.37    \\
V~I   & 6251.84 & 0.29 & $-$1.34 &9.92    &2.45    &11.06   &2.67    &$\cdots$&$\cdots$&2.64    &2.38    &3.73    &2.34    \\
V~I   & 6274.67 & 0.27 & $-$1.67 &9.36    &2.68    &10.14   &2.87    &4.59    &3.16    &2.09    &2.57    &4.19    &2.71    \\
V~I   & 6285.17 & 0.28 & $-$1.51 &7.95    &2.37    &8.82    &2.58    &$\cdots$&$\cdots$&$\cdots$&$\cdots$&$\cdots$&$\cdots$\\
V~I   & 6292.83 & 0.29 & $-$1.47 &9.39    &2.51    &7.23    &2.38    &$\cdots$&$\cdots$&$\cdots$&$\cdots$&$\cdots$&$\cdots$\\
Cr~I  & 4942.48 & 0.94 & $-$2.29 &11.12   &4.07    &$\cdots$&$\cdots$&12.12   &5.09    &$\cdots$&$\cdots$&$\cdots$&$\cdots$\\
Cr~I  & 4964.93 & 0.94 & $-$2.53 &9.43    &4.02    &10.35   &4.21    &8.33    &4.69    &$\cdots$&$\cdots$&$\cdots$&$\cdots$\\
Cr~I  & 5247.58 & 0.96 & $-$1.64 &17.33   &4.44    &15.52   &4.08    &11.67   &4.33    &13.94   &4.52    &12.72   &4.14    \\
Cr~I  & 5296.70 & 0.98 & $-$1.40 &$\cdots$&$\cdots$&18.15   &4.28    &17.82   &5.16    &$\cdots$&$\cdots$&14.27   &4.17    \\
Cr~I  & 5300.76 & 0.98 & $-$2.12 &13.78   &4.30    &10.2    &3.78    &$\cdots$&$\cdots$&$\cdots$&$\cdots$&$\cdots$&$\cdots$\\
Cr~I  & 5329.17 & 2.91 & $-$0.06 &9.43    &4.36    &7.03    &4.01    &5.19    &4.21    &$\cdots$&$\cdots$&5.73    &4.09    \\
Cr~I  & 5345.80 & 1.00 & $-$0.98 &18.16   &3.94    &$\cdots$&$\cdots$&13.12   &3.95    &15.82   &4.20    &17.91   &4.35    \\
Cr~I  & 5348.33 & 1.00 & $-$1.29 &16.78   &4.00    &17.75   &4.11    &13.10   &4.26    &13.58   &4.14    &12.50   &3.79    \\
Cr~I  & 6330.10 & 0.94 & $-$2.92 &8.09    &4.03    &9.63    &4.29    &$\cdots$&$\cdots$&3.55    &4.17    &4.72    &4.14    \\
Mn~I  & 6013.50 & 3.07 & $-$0.25 &syn     &3.59    &syn     &3.64    &syn     &3.89    &syn     &3.94    &syn     &3.79    \\
Mn~I  & 6016.65 & 3.08 & $-$0.22 &syn     &3.69    &syn     &3.49    &syn     &3.89    &syn     &3.79    &syn     &3.59    \\
Mn~I  & 6021.80 & 3.08 &    0.03 &syn     &3.64    &syn     &3.49    &syn     &3.74    &$\cdots$&$\cdots$&$\cdots$&$\cdots$\\
Fe~I  & 5060.09 & 0.00 & $-$5.46 &14.79   &5.91    &17.32   &6.31    &$\cdots$&$\cdots$&8.25    &5.60    &13.29   &6.12    \\
Fe~I  & 5068.78 & 2.94 & $-$1.04 &15.11   &5.89    &13.92   &5.57    &13.92   &6.08    &11.32   &5.47    &$\cdots$&$\cdots$\\
Fe~I  & 5131.47 & 2.22 & $-$2.52 &15.63   &6.37    &$\cdots$&$\cdots$&11.58   &6.21    &$\cdots$&$\cdots$&13.43   &6.24    \\
Fe~I  & 5159.08 & 4.28 & $-$0.82 &8.37    &6.29    &6.14    &5.92    &$\cdots$&$\cdots$&$\cdots$&$\cdots$&7.79    &6.32    \\
Fe~I  & 5162.29 & 4.18 &    0.02 &11.71   &5.91    &11.75   &5.83    &$\cdots$&$\cdots$&$\cdots$&$\cdots$&15.81   &6.69    \\
Fe~I  & 5216.29 & 1.61 & $-$2.15 &$\cdots$&$\cdots$&$\cdots$&$\cdots$&17.80   &6.17    &$\cdots$&$\cdots$&$\cdots$&$\cdots$\\
Fe~I  & 5217.40 & 3.21 & $-$1.07 &13.70   &6.01    &14.01   &5.96    &15.01   &6.62    &11.66   &5.88    &$\cdots$&$\cdots$\\
Fe~I  & 5242.52 & 3.62 & $-$0.97 &10.04   &5.82    &14.32   &6.52    &$\cdots$&$\cdots$&$\cdots$&$\cdots$&6.63    &5.46    \\
Fe~I  & 5281.79 & 3.04 & $-$0.83 &15.97   &5.90    &16.3    &5.85    &14.54   &6.06    &16.47   &6.22    &14.30   &5.78    \\
Fe~I  & 5288.55 & 3.69 & $-$1.51 &5.70    &5.73    &5.24    &5.67    &$\cdots$&$\cdots$&$\cdots$&$\cdots$&$\cdots$&$\cdots$\\
Fe~I  & 5307.38 & 1.61 & $-$2.99 &15.62   &5.94    &15.53   &5.83    &15.07   &6.49    &16.17   &6.47    &11.70   &5.56    \\
Fe~I  & 5339.95 & 3.27 & $-$0.72 &12.55   &5.49    &15.05   &5.84    &12.28   &5.83    &13.28   &5.85    &$\cdots$&$\cdots$\\
Fe~I  & 5379.57 & 3.68 & $-$1.51 &7.71    &6.04    &5.88    &5.76    &9.49    &6.69    &$\cdots$&$\cdots$&$\cdots$&$\cdots$\\
Fe~I  & 5383.39 & 4.31 &    0.65 &12.89   &5.65    &14.74   &5.86    &$\cdots$&$\cdots$&$\cdots$&$\cdots$&$\cdots$&$\cdots$\\
Fe~I  & 5569.63 & 3.42 & $-$0.49 &14.19   &5.72    &17.25   &6.12    &14.40   &6.12    &12.67   &5.67    &12.68   &5.61    \\
Fe~I  & 5586.77 & 3.37 & $-$0.12 &17.64   &5.85    &17.75   &5.76    &16.99   &6.11    &$\cdots$&$\cdots$&$\cdots$&$\cdots$\\
Fe~I  & 5859.59 & 4.55 & $-$0.42 &6.30    &5.86    &5.62    &5.75    &5.97    &6.09    &$\cdots$&$\cdots$&$\cdots$&$\cdots$\\
Fe~I  & 5862.37 & 4.55 & $-$0.13 &7.76    &5.79    &9.19    &5.98    &9.88    &6.38    &10.54   &6.37    &10.51   &6.25    \\
Fe~I  & 5916.26 & 2.45 & $-$2.83 &10.38   &5.95    &7.32    &5.52    &7.38    &6.09    &$\cdots$&$\cdots$&9.64    &6.13    \\
Fe~I  & 5927.79 & 4.65 & $-$1.09 &3.78    &6.27    &4.93    &6.46    &$\cdots$&$\cdots$&$\cdots$&$\cdots$&$\cdots$&$\cdots$\\
Fe~I  & 5934.67 & 3.93 & $-$1.17 &8.67    &6.12    &9.53    &6.21    &6.38    &6.14    &9.94    &6.54    &4.70    &5.72    \\
Fe~I  & 5956.71 & 0.86 & $-$4.61 &16.82   &6.38    &18.49   &6.57    &12.80   &6.60    &$\cdots$&$\cdots$&9.25    &5.69    \\
Fe~I  & 5976.79 & 3.94 & $-$1.31 &7.60    &6.11    &9.21    &6.32    &6.82    &6.35    &6.69    &6.24    &9.42    &6.54    \\
Fe~I  & 6003.02 & 3.88 & $-$1.12 &8.38    &5.95    &9.87    &6.14    &7.28    &6.15    &$\cdots$&$\cdots$&$\cdots$&$\cdots$\\
Fe~I  & 6008.55 & 3.88 & $-$0.99 &9.94    &6.07    &9.51    &5.96    &7.78    &6.09    &$\cdots$&$\cdots$&$\cdots$&$\cdots$\\
Fe~I  & 6024.06 & 4.55 & $-$0.12 &9.79    &6.10    &11.16   &6.26    &9.47    &6.29    &$\cdots$&$\cdots$&$\cdots$&$\cdots$\\
Fe~I  & 6027.05 & 4.08 & $-$1.09 &6.86    &5.95    &5.84    &5.81    &7.18    &6.35    &$\cdots$&$\cdots$&9.21    &6.46    \\
Fe~I  & 6056.02 & 4.73 & $-$0.46 &4.79    &5.91    &6.59    &6.17    &6.03    &6.36    &5.14    &6.16    &5.58    &6.16    \\
Fe~I  & 6065.49 & 2.61 & $-$1.53 &16.04   &5.76    &17.47   &5.89    &15.33   &6.19    &12.99   &5.64    &16.10   &6.02    \\
Fe~I  & 6082.72 & 2.22 & $-$3.57 &7.56    &5.92    &7.3     &5.92    &2.81    &5.86    &$\cdots$&$\cdots$&$\cdots$&$\cdots$\\
Fe~I  & 6151.62 & 2.18 & $-$3.30 &10.17   &5.94    &10.19   &5.94    &7.49    &6.20    &$\cdots$&$\cdots$&$\cdots$&$\cdots$\\
Fe~I  & 6165.36 & 4.14 & $-$1.47 &4.05    &5.99    &4.8     &6.12    &$\cdots$&$\cdots$&$\cdots$&$\cdots$&$\cdots$&$\cdots$\\
Fe~I  & 6173.33 & 2.22 & $-$2.88 &11.75   &5.83    &11.32   &5.74    &12.09   &6.50    &8.88    &5.88    &11.95   &6.17    \\
Fe~I  & 6180.21 & 2.73 & $-$2.59 &9.66    &5.95    &9.43    &5.91    &5.13    &5.87    &6.37    &5.92    &$\cdots$&$\cdots$\\
Fe~I  & 6219.29 & 2.20 & $-$2.43 &14.92   &5.82    &15.14   &5.80    &11.60   &5.94    &10.64   &5.64    &10.20   &5.45    \\
Fe~I  & 6229.25 & 2.83 & $-$2.81 &6.75    &5.91    &5.24    &5.74    &5.96    &6.35    &7.77    &6.47    &$\cdots$&$\cdots$\\
Fe~I  & 6240.67 & 2.22 & $-$3.23 &10.35   &5.97    &9.97    &5.91    &7.30    &6.17    &$\cdots$&$\cdots$&$\cdots$&$\cdots$\\
Fe~I  & 6246.34 & 3.60 & $-$0.73 &12.45   &5.80    &13.65   &5.91    &11.16   &5.97    &10.21   &5.70    &$\cdots$&$\cdots$\\
Fe~I  & 6252.58 & 2.40 & $-$1.69 &18.47   &5.94    &18.25   &5.82    &16.10   &6.17    &15.41   &5.88    &14.79   &5.66    \\
Fe~I  & 6254.27 & 2.28 & $-$2.44 &16.75   &6.29    &18.27   &6.44    &14.87   &6.60    &12.53   &6.04    &$\cdots$&$\cdots$\\
Fe~I  & 6265.15 & 2.18 & $-$2.55 &16.26   &6.09    &17.7    &6.23    &12.20   &6.11    &14.07   &6.22    &$\cdots$&$\cdots$\\
Fe~I  & 6270.24 & 2.86 & $-$2.46 &7.91    &5.75    &11.36   &6.23    &5.98    &6.02    &7.16    &6.06    &$\cdots$&$\cdots$\\
Fe~I  & 6301.51 & 3.65 & $-$0.72 &12.24   &5.81    &14.54   &6.10    &11.20   &6.01    &7.86    &5.42    &$\cdots$&$\cdots$\\
Fe~I  & 6322.69 & 2.59 & $-$2.43 &12.74   &6.05    &13.78   &6.16    &10.52   &6.26    &14.18   &6.66    &11.54   &6.14    \\
Fe~I  & 6344.15 & 2.43 & $-$2.92 &13.43   &6.42    &14.95   &6.60    &$\cdots$&$\cdots$&$\cdots$&$\cdots$&12.24   &6.53    \\
Fe~I  & 6355.03 & 2.84 & $-$2.35 &11.94   &6.21    &10.03   &5.90    &7.52    &6.09    &8.7     &6.12    &8.35    &5.95    \\
Fe~I  & 6358.71 & 0.86 & $-$4.47 &16.83   &6.12    &16.04   &5.96    &13.74   &6.53    &6.66    &5.41    &16.03   &6.44    \\
Fe~I  & 6380.77 & 4.19 & $-$1.38 &4.63    &6.04    &6.96    &6.38    &$\cdots$&$\cdots$&$\cdots$&$\cdots$&$\cdots$&$\cdots$\\
Fe~I  & 6393.62 & 2.43 & $-$1.43 &$\cdots$&$\cdots$&$\cdots$&$\cdots$&17.80   &6.18    &$\cdots$&$\cdots$&$\cdots$&$\cdots$\\
Fe~I  & 6411.66 & 3.65 & $-$0.60 &13.67   &5.90    &12.37   &5.63    &10.83   &5.82    &$\cdots$&$\cdots$&$\cdots$&$\cdots$\\
Fe~I  & 6419.97 & 4.73 & $-$0.24 &$\cdots$&$\cdots$&$\cdots$&$\cdots$&$\cdots$&$\cdots$&$\cdots$&$\cdots$&$\cdots$&$\cdots$\\
Fe~I  & 6421.37 & 2.28 & $-$2.03 &18.27   &6.08    &17.96   &5.93    &17.01   &6.49    &16.81   &6.26    &$\cdots$&$\cdots$\\
Fe~I  & 6430.86 & 2.18 & $-$2.01 &$\cdots$&$\cdots$&$\cdots$&$\cdots$&16.54   &6.20    &$\cdots$&$\cdots$&$\cdots$&$\cdots$\\
Fe~I  & 6475.63 & 2.52 & $-$2.94 &10.60   &6.17    &8.48    &5.88    &10.12   &6.67    &7.44    &6.18    &7.58    &6.05    \\
Fe~I  & 6518.37 & 2.83 & $-$2.46 &9.92    &5.97    &10.95   &6.10    &8.75    &6.34    &8.38    &6.16    &6.06    &5.72    \\
Fe~I  & 6546.25 & 2.76 & $-$1.54 &15.33   &5.77    &14.29   &5.55    &14.00   &6.10    &$\cdots$&$\cdots$&$\cdots$&$\cdots$\\
Fe~I  & 6569.23 & 4.73 & $-$0.42 &8.02    &6.33    &6.84    &6.14    &6.66    &6.39    &$\cdots$&$\cdots$&$\cdots$&$\cdots$\\
Fe~I  & 6581.21 & 1.48 & $-$4.68 &6.74    &5.81    &7.27    &5.93    &$\cdots$&$\cdots$&$\cdots$&$\cdots$&8.11    &6.41    \\
Fe~I  & 6593.88 & 2.43 & $-$2.42 &14.96   &6.11    &14.63   &6.01    &12.83   &6.38    &14.25   &6.43    &$\cdots$&$\cdots$\\
Fe~I  & 6608.04 & 2.28 & $-$4.03 &5.90    &6.19    &7.11    &6.38    &$\cdots$&$\cdots$&$\cdots$&$\cdots$&$\cdots$&$\cdots$\\
Fe~I  & 6677.98 & 2.69 & $-$1.42 &17.66   &5.87    &18.58   &5.91    &$\cdots$&$\cdots$&14.08   &5.73    &$\cdots$&$\cdots$\\
Fe~I  & 6703.56 & 2.76 & $-$3.16 &5.94    &6.00    &5.05    &5.92    &6.23    &6.60    &$\cdots$&$\cdots$&$\cdots$&$\cdots$\\
Fe~I  & 6710.34 & 1.49 & $-$4.88 &5.90    &5.90    &3.87    &5.68    &$\cdots$&$\cdots$&$\cdots$&$\cdots$&$\cdots$&$\cdots$\\
Fe~I  & 6750.18 & 2.42 & $-$2.62 &12.41   &5.90    &15.85   &6.36    &12.65   &6.53    &12.77   &6.38    &$\cdots$&$\cdots$\\
Fe~II & 4993.36 & 2.81 & $-$3.62 &4.10    &6.05    &4.46    &6.07    &$\cdots$&$\cdots$&$\cdots$&$\cdots$&$\cdots$&$\cdots$\\
Fe~II & 5234.65 & 3.22 & $-$2.18 &10.03   &6.27    &9.41    &6.03    &10.72   &6.19    &12.05   &6.32    &10.96   &6.24    \\
Fe~II & 5425.25 & 3.20 & $-$3.22 &5.83    &6.50    &3.12    &5.90    &7.01    &6.59    &$\cdots$&$\cdots$&4.81    &6.35    \\
Fe~II & 6247.60 & 3.89 & $-$2.30 &3.86    &6.09    &3.79    &6.01    &6.14    &6.35    &$\cdots$&$\cdots$&4.21    &6.04    \\
Fe~II & 6432.68 & 2.89 & $-$3.57 &4.63    &6.21    &7.14    &6.57    &7.37    &6.59    &$\cdots$&$\cdots$&5.32    &6.37    \\
Fe~II & 6456.44 & 3.90 & $-$2.05 &$\cdots$&$\cdots$&$\cdots$&$\cdots$&$\cdots$&$\cdots$&$\cdots$&$\cdots$&$\cdots$&$\cdots$\\
Fe~II & 6516.10 & 2.89 & $-$3.31 &4.93    &6.01    &5.74    &6.09    &6.17    &6.15    &$\cdots$&$\cdots$&7.10    &6.31    \\
Co~I  & 5301.04 & 1.71 & $-$2.00 &8.49    &3.81    &6.89    &3.60    &$\cdots$&$\cdots$&$\cdots$&$\cdots$&$\cdots$&$\cdots$\\
Co~I  & 5647.23 & 2.28 & $-$1.56 &4.44    &3.54    &3.31    &3.41    &$\cdots$&$\cdots$&$\cdots$&$\cdots$&$\cdots$&$\cdots$\\
Co~I  & 5991.86 & 2.08 & $-$1.85 &4.55    &3.54    &4.71    &3.61    &$\cdots$&$\cdots$&$\cdots$&$\cdots$&$\cdots$&$\cdots$\\
Co~I  & 6771.05 & 1.88 & $-$1.97 &6.69    &3.62    &4.76    &3.41    &$\cdots$&$\cdots$&$\cdots$&$\cdots$&$\cdots$&$\cdots$\\
Ni~I  & 4904.42 & 3.54 & $-$0.17 &8.70    &4.61    &13.25   &5.36    &$\cdots$&$\cdots$&$\cdots$&$\cdots$&$\cdots$&$\cdots$\\
Ni~I  & 5035.39 & 3.63 &    0.29 &11.30   &4.75    &12.09   &4.79    &7.97    &4.38    &$\cdots$&$\cdots$&10.70   &4.67    \\
Ni~I  & 5080.53 & 3.65 &    0.13 &9.75    &4.62    &13.8    &5.27    &$\cdots$&$\cdots$&6.11    &4.20    &10.79   &4.86    \\
Ni~I  & 5084.11 & 3.68 &    0.03 &7.83    &4.42    &7.05    &4.26    &$\cdots$&$\cdots$&$\cdots$&$\cdots$&$\cdots$&$\cdots$\\
Ni~I  & 5146.48 & 3.71 & $-$0.06 &9.48    &4.83    &11.37   &5.09    &12.41   &5.56    &$\cdots$&$\cdots$&10.74   &5.11    \\
Ni~I  & 5578.73 & 1.68 & $-$2.64 &12.99   &5.06    &13.39   &5.06    &12.22   &5.47    &$\cdots$&$\cdots$&11.29   &5.03    \\
Ni~I  & 5587.85 & 1.94 & $-$2.14 &$\cdots$&$\cdots$&$\cdots$&$\cdots$&9.10    &4.80    &$\cdots$&$\cdots$&10.63   &4.77    \\
Ni~I  & 5592.29 & 1.95 & $-$2.59 &$\cdots$&$\cdots$&$\cdots$&$\cdots$&$\cdots$&$\cdots$&$\cdots$&$\cdots$&4.48    &4.37    \\
Ni~I  & 5846.99 & 1.68 & $-$3.21 &5.01    &4.36    &4.68    &4.36    &$\cdots$&$\cdots$&$\cdots$&$\cdots$&$\cdots$&$\cdots$\\
Ni~I  & 6108.12 & 1.68 & $-$2.45 &11.38   &4.50    &12.32   &4.61    &8.61    &4.66    &6.38    &4.24    &$\cdots$&$\cdots$\\
Ni~I  & 6128.98 & 1.68 & $-$3.33 &6.48    &4.67    &4.49    &4.43    &$\cdots$&$\cdots$&$\cdots$&$\cdots$&6.70    &5.07    \\
Ni~I  & 6176.80 & 4.09 & $-$0.43 &5.05    &4.91    &4.92    &4.88    &6.58    &5.38    &$\cdots$&$\cdots$&$\cdots$&$\cdots$\\
Ni~I  & 6177.25 & 1.83 & $-$3.60 &3.46    &4.70    &3.73    &4.79    &$\cdots$&$\cdots$&$\cdots$&$\cdots$&$\cdots$&$\cdots$\\
Ni~I  & 6327.59 & 1.68 & $-$3.15 &8.92    &4.80    &11.02   &5.09    &4.58    &4.81    &$\cdots$&$\cdots$&5.80    &4.64    \\
Ni~I  & 6482.79 & 1.94 & $-$2.63 &8.93    &4.64    &8.76    &4.62    &8.09    &5.07    &6.05    &4.68    &9.15    &4.94    \\
Ni~I  & 6532.90 & 1.94 & $-$3.39 &4.03    &4.71    &3.43    &4.66    &$\cdots$&$\cdots$&$\cdots$&$\cdots$&$\cdots$&$\cdots$\\
Ni~I  & 6586.32 & 1.95 & $-$2.81 &7.82    &4.68    &8.29    &4.75    &8.41    &5.30    &2.89    &4.41    &3.92    &4.44    \\
Ni~I  & 6643.64 & 1.68 & $-$2.30 &16.92   &5.08    &18.01   &5.16    &14.48   &5.32    &10      &4.50    &$\cdots$&$\cdots$\\
Ni~I  & 6767.78 & 1.83 & $-$2.17 &13.64   &4.67    &14.38   &4.73    &12.72   &5.10    &12.83   &4.95    &10.63   &4.51    \\
Ni~I  & 6772.30 & 3.66 & $-$0.98 &5.24    &4.87    &4.96    &4.83    &$\cdots$&$\cdots$&$\cdots$&$\cdots$&$\cdots$&$\cdots$\\
Cu~I  & 5105.55 & 1.39 & $-$1.51 &syn     &2.38    &syn     &2.44    &syn     &2.35    &syn     &2.51    &syn     &2.31    \\
Y~II  & 5087.44 & 1.08 & $-$0.16 &9.53    &0.63    &12      &0.99    &10.65   &1.07    &9.3     &0.72    &8.78    &0.59    \\
Y~II  & 5200.44 & 0.99 & $-$0.57 &5.88    &0.28    &7.16    &0.47    &$\cdots$&$\cdots$&$\cdots$&$\cdots$&8.22    &0.78    \\
Y~II  & 5509.94 & 0.99 & $-$1.01 &5.82    &0.69    &6.14    &0.74    &$\cdots$&$\cdots$&5.93    &0.94    &$\cdots$&$\cdots$\\
Zr~I  & 6127.47 & 0.15 & $-$1.06 &4.20    &1.15    &3.17    &1.13    &$\cdots$&$\cdots$&$\cdots$&$\cdots$&$\cdots$&$\cdots$\\
Zr~I  & 6134.55 & 0.00 & $-$1.28 &5.72    &1.33    &3.99    &1.24    &$\cdots$&$\cdots$&$\cdots$&$\cdots$&$\cdots$&$\cdots$\\
Zr~II & 5112.32 & 1.66 & $-$0.59 &4.48    &1.46    &3.62    &1.31    &$\cdots$&$\cdots$&$\cdots$&$\cdots$&$\cdots$&$\cdots$\\
Ba~II & 5853.70 & 0.60 & $-$1.01 &13.61   &0.78    &16.31   &1.12    &13.67   &1.16    &13.82   &1.02    &12.55   &0.76    \\
Ba~II & 6141.72 & 0.70 & $-$0.08 &syn     &0.73    &17.79   &0.48    &18.33   &1.02    &16.78   &0.62    &$\cdots$&$\cdots$\\
Ba~II & 6496.92 & 0.60 & $-$0.38 &syn     &0.66    &syn     &0.49    &syn     &0.83    &18.34   &0.94    &syn     &0.73    \\
La~II & 6320.41 & 0.17 & $-$1.56 &4.45    &0.10    &4.88    &0.19    &$\cdots$&$\cdots$&$\cdots$&$\cdots$&$\cdots$&$\cdots$\\
La~II & 6390.50 & 4.15 & $-$1.40 &4.53    &0.16    &2.76    &$-$0.11 &$\cdots$&$\cdots$&$\cdots$&$\cdots$&$\cdots$&$\cdots$\\
Ce~II & 5274.26 & 1.04 &    0.15 &5.19    &0.49    &4.45    &0.39    &$\cdots$&$\cdots$&7.28    &1.05    &$\cdots$&$\cdots$\\
Nd~II & 5249.58 & 0.98 &    0.22 &9.08    &0.72    &6.49    &0.28    &$\cdots$&$\cdots$&7.81    &0.71    &8.60    &0.77    \\
Nd~II & 5319.82 & 0.55 & $-$0.19 &11.40   &0.91    &9.48    &0.53    &7.71    &0.69    &9.27    &0.79    &$\cdots$&$\cdots$\\
Eu~II & 6645.12 & 1.38 &    0.12 &syn     &$-$0.24 &syn     &$-$0.06 &$\cdots$&$\cdots$&$\cdots$&$\cdots$&$\cdots$&$\cdots$\\
Dy~II & 5169.69 & 0.10 & $-$1.66 &4.18    &0.35    &3.38    &0.25    &$\cdots$&$\cdots$&4.49    &0.76    &$\cdots$&$\cdots$\\                                                                                                                                                                                              
                                                                                                                                                                                             
\end{longtable}                                                                                                                                                                                                                                                                                                                                                                     
} 
\onllongtab{2}{
\tiny
\begin{longtable}{lccrrrrrrrrr}
\caption{\label{Tab:Linelist2} Line data, equivalent widths, and and abundances
  derived for each line for the sample of Pal~14 giants.}\\
\hline
\hline
Species& $\lambda$ & $\chi$ & $\log gf$ & EWHH201&$A$& EWHV043&$A$&EWHV104&$A$ &EWHV074&$A$ \\
&[{\AA}]&[eV]&&[pm]&&[pm]&&[pm]&&[pm]&\\
\hline
\endfirsthead
\caption{Continued.} \\
\hline
Species& $\lambda$ & $\chi$ & $\log gf$ & EWHH201&$A$& EWHV043&$A$&EWHV104&$A$ &EWHV074&$A$ \\
&[{\AA}]&[eV]&&[pm]&&[pm]&&[pm]&&[pm]&\\

\hline
\endhead
\hline
\endfoot
\hline
\endlastfoot
O~I   & 6300.33 & 0.00 & $-$9.82 &$\cdots$&$\cdots$&$\cdots$&$\cdots$&$\cdots$&$\cdots$&$\cdots$&$\cdots$\\
O~I   & 6363.76 & 0.02 & $-$10.30&$\cdots$&$\cdots$&$\cdots$&$\cdots$&$\cdots$&$\cdots$&$\cdots$&$\cdots$\\
Na~I  & 5682.64 & 2.10 & $-$0.70 &$\cdots$&$\cdots$&$\cdots$&$\cdots$&$\cdots$&$\cdots$&$\cdots$&$\cdots$\\
Na~I  & 5688.21 & 2.10 & $-$0.42 &$\cdots$&$\cdots$&$\cdots$&$\cdots$&$\cdots$&$\cdots$&$\cdots$&$\cdots$\\
Mg~I  & 5528.42 & 4.35 & $-$0.36 &17.18   &6.02    &syn     &6.68    &syn     &6.64    &syn     &6.48    \\
Mg~I  & 5711.11 & 4.33 & $-$1.73 &10.44   &6.49    &11.46   &6.89    &14.21   &7.12    &9.59    &6.66    \\
Si~I  & 5690.47 & 4.93 & $-$1.77 &5.13    &6.68    &$\cdots$&$\cdots$&$\cdots$&$\cdots$&$\cdots$&$\cdots$\\
Si~I  & 5948.55 & 5.08 & $-$0.78 &$\cdots$&$\cdots$&$\cdots$&$\cdots$&$\cdots$&$\cdots$&10.97   &6.79    \\
Si~I  & 6155.16 & 5.61 & $-$0.75 &$\cdots$&$\cdots$&$\cdots$&$\cdots$&$\cdots$&$\cdots$&6.98    &6.84    \\
Ca~I  & 5261.71 & 2.52 & $-$0.58 &10.29   &5.06    &$\cdots$&$\cdots$&$\cdots$&$\cdots$&$\cdots$&$\cdots$\\
Ca~I  & 5581.97 & 2.53 & $-$0.56 &$\cdots$&$\cdots$&$\cdots$&$\cdots$&13.25   &5.63    &11.69   &5.59    \\
Ca~I  & 5601.28 & 2.53 & $-$0.52 &10.24   &4.97    &$\cdots$&$\cdots$&11.03   &5.27    &9.76    &5.25    \\
Ca~I  & 5857.46 & 2.93 &    0.24 &12.70   &5.06    &$\cdots$&$\cdots$&$\cdots$&$\cdots$&$\cdots$&$\cdots$\\
Ca~I  & 6122.24 & 1.89 & $-$0.32 &$\cdots$&$\cdots$&14.81   &5.02    &$\cdots$&$\cdots$&$\cdots$&$\cdots$\\
Ca~I  & 6166.45 & 2.52 & $-$1.14 &7.96    &5.24    &7.85    &5.50    &10.30   &5.73    &$\cdots$&$\cdots$\\
Ca~I  & 6169.05 & 2.52 & $-$0.80 &7.38    &4.83    &$\cdots$&$\cdots$&$\cdots$&$\cdots$&$\cdots$&$\cdots$\\
Ca~I  & 6169.56 & 2.52 & $-$0.48 &$\cdots$&$\cdots$&$\cdots$&$\cdots$&$\cdots$&$\cdots$&9.65    &5.14    \\
Ca~I  & 6439.09 & 2.53 &    0.39 &$\cdots$&$\cdots$&15.25   &5.18    &17.82   &5.27    &17.62   &5.50    \\
Ca~I  & 6455.59 & 2.89 & $-$1.29 &8.49    &5.44    &$\cdots$&$\cdots$&8.60    &5.70    &7.61    &5.65    \\
Ca~I  & 6471.67 & 2.52 & $-$0.69 &$\cdots$&$\cdots$&$\cdots$&$\cdots$&$\cdots$&$\cdots$&11.64   &5.64    \\
Ca~I  & 6499.63 & 2.52 & $-$0.82 &12.27   &5.47    &$\cdots$&$\cdots$&7.22    &5.01    &8.88    &5.35    \\
Ca~I  & 6572.78 & 0.00 & $-$4.24 &$\cdots$&$\cdots$&$\cdots$&$\cdots$&$\cdots$&$\cdots$&$\cdots$&$\cdots$\\
Ca~I  & 6717.69 & 2.71 & $-$0.52 &9.20    &5.07    &$\cdots$&$\cdots$&9.79    &5.25    &$\cdots$&$\cdots$\\
Sc~II & 5239.84 & 1.46 & $-$0.77 &$\cdots$&$\cdots$&$\cdots$&$\cdots$&$\cdots$&$\cdots$&$\cdots$&$\cdots$\\
Sc~II & 5526.83 & 1.77 &    0.02 &$\cdots$&$\cdots$&13.62   &2.38    &$\cdots$&$\cdots$&$\cdots$&$\cdots$\\
Sc~II & 5669.03 & 1.50 & $-$1.20 &7.99    &2.13    &$\cdots$&$\cdots$&$\cdots$&$\cdots$&4.48    &1.87    \\
Sc~II & 5684.22 & 1.51 & $-$1.07 &$\cdots$&$\cdots$&$\cdots$&$\cdots$&$\cdots$&$\cdots$&$\cdots$&$\cdots$\\
Sc~II & 6245.62 & 1.51 & $-$1.03 &8.81    &2.04    &8.54    &2.21    &$\cdots$&$\cdots$&5.17    &1.78    \\
Sc~II & 6604.61 & 1.36 & $-$1.31 &$\cdots$&$\cdots$&$\cdots$&$\cdots$&$\cdots$&$\cdots$&$\cdots$&$\cdots$\\
Ti~I  & 4997.10 & 0.00 & $-$2.12 &$\cdots$&$\cdots$&$\cdots$&$\cdots$&$\cdots$&$\cdots$&$\cdots$&$\cdots$\\
Ti~I  & 4999.51 & 0.83 &    0.25 &$\cdots$&$\cdots$&$\cdots$&$\cdots$&$\cdots$&$\cdots$&12.47   &3.47    \\
Ti~I  & 5009.66 & 0.02 & $-$2.26 &$\cdots$&$\cdots$&$\cdots$&$\cdots$&$\cdots$&$\cdots$&$\cdots$&$\cdots$\\
Ti~I  & 5022.89 & 0.83 & $-$0.43 &$\cdots$&$\cdots$&12.77   &4.17    &$\cdots$&$\cdots$&7.86    &3.41    \\
Ti~I  & 5045.43 & 0.85 & $-$2.00 &$\cdots$&$\cdots$&$\cdots$&$\cdots$&$\cdots$&$\cdots$&$\cdots$&$\cdots$\\
Ti~I  & 5062.13 & 2.16 & $-$0.46 &$\cdots$&$\cdots$&$\cdots$&$\cdots$&$\cdots$&$\cdots$&$\cdots$&$\cdots$\\
Ti~I  & 5147.49 & 0.00 & $-$1.88 &9.55    &3.41    &12.50   &4.60    &$\cdots$&$\cdots$&$\cdots$&$\cdots$\\
Ti~I  & 5152.18 & 0.02 & $-$1.91 &11.56   &3.73    &$\cdots$&$\cdots$&$\cdots$&$\cdots$&$\cdots$&$\cdots$\\
Ti~I  & 5210.39 & 0.05 & $-$0.88 &17.37   &3.57    &$\cdots$&$\cdots$&$\cdots$&$\cdots$&11.48   &3.42    \\
Ti~I  & 5219.71 & 0.02 & $-$1.98 &$\cdots$&$\cdots$&$\cdots$&$\cdots$&$\cdots$&$\cdots$&7.68    &4.22    \\
Ti~I  & 5401.35 & 0.82 & $-$2.89 &$\cdots$&$\cdots$&$\cdots$&$\cdots$&$\cdots$&$\cdots$&$\cdots$&$\cdots$\\
Ti~I  & 5644.13 & 2.27 &    0.21 &$\cdots$&$\cdots$&$\cdots$&$\cdots$&$\cdots$&$\cdots$&$\cdots$&$\cdots$\\
Ti~I  & 5648.54 & 2.50 & $-$0.26 &$\cdots$&$\cdots$&$\cdots$&$\cdots$&$\cdots$&$\cdots$&$\cdots$&$\cdots$\\
Ti~I  & 5866.46 & 1.07 & $-$0.84 &$\cdots$&$\cdots$&10.42   &4.31    &$\cdots$&$\cdots$&$\cdots$&$\cdots$\\
Ti~I  & 5918.55 & 1.07 & $-$1.46 &$\cdots$&$\cdots$&$\cdots$&$\cdots$&$\cdots$&$\cdots$&$\cdots$&$\cdots$\\
Ti~I  & 5922.11 & 1.05 & $-$1.47 &3.61    &3.57    &$\cdots$&$\cdots$&$\cdots$&$\cdots$&$\cdots$&$\cdots$\\
Ti~I  & 5941.76 & 1.05 & $-$1.51 &$\cdots$&$\cdots$&$\cdots$&$\cdots$&$\cdots$&$\cdots$&$\cdots$&$\cdots$\\
Ti~I  & 5965.82 & 1.88 & $-$0.41 &$\cdots$&$\cdots$&$\cdots$&$\cdots$&$\cdots$&$\cdots$&$\cdots$&$\cdots$\\
Ti~I  & 5978.54 & 1.88 & $-$0.50 &$\cdots$&$\cdots$&$\cdots$&$\cdots$&$\cdots$&$\cdots$&$\cdots$&$\cdots$\\
Ti~I  & 6064.62 & 1.05 & $-$1.94 &$\cdots$&$\cdots$&$\cdots$&$\cdots$&$\cdots$&$\cdots$&$\cdots$&$\cdots$\\
Ti~I  & 6126.22 & 1.07 & $-$1.42 &6.78    &3.95    &7.41    &4.41    &$\cdots$&$\cdots$&7.6     &4.55    \\
Ti~I  & 6258.12 & 1.44 & $-$0.36 &$\cdots$&$\cdots$&$\cdots$&$\cdots$&10.75   &4.17    &4.72    &3.56    \\
Ti~I  & 6303.77 & 1.44 & $-$1.57 &$\cdots$&$\cdots$&$\cdots$&$\cdots$&$\cdots$&$\cdots$&$\cdots$&$\cdots$\\
Ti~I  & 6554.25 & 1.44 & $-$1.22 &$\cdots$&$\cdots$&$\cdots$&$\cdots$&$\cdots$&$\cdots$&$\cdots$&$\cdots$\\
Ti~I  & 6556.07 & 1.46 & $-$1.07 &6.59    &3.94    &$\cdots$&$\cdots$&7.17    &4.46    &$\cdots$&$\cdots$\\
Ti~I  & 6599.10 & 0.90 & $-$2.09 &$\cdots$&$\cdots$&5.60    &4.58    &7.00    &4.75    &$\cdots$&$\cdots$\\
Ti~I  & 6743.15 & 0.90 & $-$1.63 &$\cdots$&$\cdots$&$\cdots$&$\cdots$&4.97    &4.04    &4.76    &4.14    \\
Ti~II & 5005.16 & 1.57 & $-$2.72 &7.25    &3.91    &$\cdots$&$\cdots$&$\cdots$&$\cdots$&$\cdots$&$\cdots$\\
Ti~II & 5013.70 & 1.58 & $-$2.19 &$\cdots$&$\cdots$&$\cdots$&$\cdots$&$\cdots$&$\cdots$&$\cdots$&$\cdots$\\
Ti~II & 5185.91 & 1.89 & $-$1.35 &$\cdots$&$\cdots$&12.80   &4.23    &$\cdots$&$\cdots$&11.78   &4.15    \\
Ti~II & 5226.54 & 1.57 & $-$1.26 &$\cdots$&$\cdots$&16.76   &4.42    &$\cdots$&$\cdots$&$\cdots$&$\cdots$\\
Ti~II & 5336.81 & 1.58 & $-$1.70 &$\cdots$&$\cdots$&$\cdots$&$\cdots$&$\cdots$&$\cdots$&$\cdots$&$\cdots$\\
Ti~II & 5396.27 & 1.58 & $-$2.51 &$\cdots$&$\cdots$&$\cdots$&$\cdots$&$\cdots$&$\cdots$&$\cdots$&$\cdots$\\
Ti~II & 5418.79 & 1.58 & $-$2.00 &10.70   &3.82    &8.85    &3.78    &10.26   &3.87    &9.4     &3.87    \\
Ti~II & 6559.55 & 2.05 & $-$2.02 &$\cdots$&$\cdots$&5.39    &3.78    &3.91    &3.55    &6.64    &4.00    \\
V~I   & 5627.64 & 1.08 & $-$0.36 &$\cdots$&$\cdots$&$\cdots$&$\cdots$&$\cdots$&$\cdots$&$\cdots$&$\cdots$\\
V~I   & 5670.85 & 1.08 & $-$0.42 &$\cdots$&$\cdots$&$\cdots$&$\cdots$&$\cdots$&$\cdots$&$\cdots$&$\cdots$\\
V~I   & 5703.58 & 1.05 & $-$0.21 &$\cdots$&$\cdots$&$\cdots$&$\cdots$&$\cdots$&$\cdots$&$\cdots$&$\cdots$\\
V~I   & 6039.73 & 1.06 & $-$0.65 &3.99    &2.81    &$\cdots$&$\cdots$&$\cdots$&$\cdots$&$\cdots$&$\cdots$\\
V~I   & 6081.47 & 1.05 & $-$0.58 &$\cdots$&$\cdots$&$\cdots$&$\cdots$&$\cdots$&$\cdots$&$\cdots$&$\cdots$\\
V~I   & 6090.21 & 1.08 & $-$0.06 &$\cdots$&$\cdots$&$\cdots$&$\cdots$&$\cdots$&$\cdots$&$\cdots$&$\cdots$\\
V~I   & 6111.66 & 1.04 & $-$0.72 &$\cdots$&$\cdots$&$\cdots$&$\cdots$&$\cdots$&$\cdots$&$\cdots$&$\cdots$\\
V~I   & 6135.38 & 1.05 & $-$0.75 &$\cdots$&$\cdots$&$\cdots$&$\cdots$&$\cdots$&$\cdots$&$\cdots$&$\cdots$\\
V~I   & 6216.37 & 4.73 & $-$1.42 &$\cdots$&$\cdots$&$\cdots$&$\cdots$&$\cdots$&$\cdots$&$\cdots$&$\cdots$\\
V~I   & 6224.51 & 0.29 & $-$2.01 &$\cdots$&$\cdots$&$\cdots$&$\cdots$&$\cdots$&$\cdots$&$\cdots$&$\cdots$\\
V~I   & 6243.11 & 0.30 & $-$0.98 &6.25    &2.40    &$\cdots$&$\cdots$&$\cdots$&$\cdots$&$\cdots$&$\cdots$\\
V~I   & 6251.84 & 0.29 & $-$1.34 &$\cdots$&$\cdots$&$\cdots$&$\cdots$&$\cdots$&$\cdots$&$\cdots$&$\cdots$\\
V~I   & 6274.67 & 0.27 & $-$1.67 &4.16    &2.78    &$\cdots$&$\cdots$&$\cdots$&$\cdots$&$\cdots$&$\cdots$\\
V~I   & 6285.17 & 0.28 & $-$1.51 &$\cdots$&$\cdots$&$\cdots$&$\cdots$&$\cdots$&$\cdots$&$\cdots$&$\cdots$\\
V~I   & 6292.83 & 0.29 & $-$1.47 &$\cdots$&$\cdots$&$\cdots$&$\cdots$&$\cdots$&$\cdots$&$\cdots$&$\cdots$\\
Cr~I  & 4942.48 & 0.94 & $-$2.29 &$\cdots$&$\cdots$&$\cdots$&$\cdots$&$\cdots$&$\cdots$&$\cdots$&$\cdots$\\
Cr~I  & 4964.93 & 0.94 & $-$2.53 &$\cdots$&$\cdots$&$\cdots$&$\cdots$&$\cdots$&$\cdots$&$\cdots$&$\cdots$\\
Cr~I  & 5247.58 & 0.96 & $-$1.64 &$\cdots$&$\cdots$&$\cdots$&$\cdots$&$\cdots$&$\cdots$&$\cdots$&$\cdots$\\
Cr~I  & 5296.70 & 0.98 & $-$1.40 &17.25   &4.59    &$\cdots$&$\cdots$&$\cdots$&$\cdots$&$\cdots$&$\cdots$\\
Cr~I  & 5300.76 & 0.98 & $-$2.12 &9.69    &4.17    &$\cdots$&$\cdots$&$\cdots$&$\cdots$&$\cdots$&$\cdots$\\
Cr~I  & 5329.17 & 2.91 & $-$0.06 &5.46    &4.07    &$\cdots$&$\cdots$&$\cdots$&$\cdots$&$\cdots$&$\cdots$\\
Cr~I  & 5345.80 & 1.00 & $-$0.98 &15.99   &3.99    &$\cdots$&$\cdots$&13.57   &3.95    &13.42   &4.22    \\
Cr~I  & 5348.33 & 1.00 & $-$1.29 &12.68   &3.79    &$\cdots$&$\cdots$&10.42   &3.79    &12.75   &4.41    \\
Cr~I  & 6330.10 & 0.94 & $-$2.92 &7.58    &4.55    &$\cdots$&$\cdots$&$\cdots$&$\cdots$&$\cdots$&$\cdots$\\
Mn~I  & 6013.50 & 3.07 & $-$0.25 &$\cdots$&$\cdots$&$\cdots$&$\cdots$&$\cdots$&$\cdots$&$\cdots$&$\cdots$\\
Mn~I  & 6016.65 & 3.08 & $-$0.22 &$\cdots$&$\cdots$&$\cdots$&$\cdots$&$\cdots$&$\cdots$&$\cdots$&$\cdots$\\
Mn~I  & 6021.80 & 3.08 &    0.03 &$\cdots$&$\cdots$&$\cdots$&$\cdots$&$\cdots$&$\cdots$&$\cdots$&$\cdots$\\
Fe~I  & 5060.09 & 0.00 & $-$5.46 &9.33    &5.54    &$\cdots$&$\cdots$&$\cdots$&$\cdots$&$\cdots$&$\cdots$\\
Fe~I  & 5068.78 & 2.94 & $-$1.04 &17.59   &6.36    &12.04   &5.91    &$\cdots$&$\cdots$&$\cdots$&$\cdots$\\
Fe~I  & 5131.47 & 2.22 & $-$2.52 &11.12   &5.81    &$\cdots$&$\cdots$&$\cdots$&$\cdots$&5.9     &5.53    \\
Fe~I  & 5159.08 & 4.28 & $-$0.82 &$\cdots$&$\cdots$&$\cdots$&$\cdots$&$\cdots$&$\cdots$&$\cdots$&$\cdots$\\
Fe~I  & 5162.29 & 4.18 &    0.02 &11.87   &5.94    &11.94   &6.34    &$\cdots$&$\cdots$&$\cdots$&$\cdots$\\
Fe~I  & 5216.29 & 1.61 & $-$2.15 &$\cdots$&$\cdots$&$\cdots$&$\cdots$&$\cdots$&$\cdots$&$\cdots$&$\cdots$\\
Fe~I  & 5217.40 & 3.21 & $-$1.07 &10.49   &5.55    &9.60    &5.79    &$\cdots$&$\cdots$&$\cdots$&$\cdots$\\
Fe~I  & 5242.52 & 3.62 & $-$0.97 &12.14   &6.27    &$\cdots$&$\cdots$&$\cdots$&$\cdots$&$\cdots$&$\cdots$\\
Fe~I  & 5281.79 & 3.04 & $-$0.83 &13.87   &5.62    &$\cdots$&$\cdots$&$\cdots$&$\cdots$&$\cdots$&$\cdots$\\
Fe~I  & 5288.55 & 3.69 & $-$1.51 &$\cdots$&$\cdots$&$\cdots$&$\cdots$&$\cdots$&$\cdots$&$\cdots$&$\cdots$\\
Fe~I  & 5307.38 & 1.61 & $-$2.99 &15.83   &6.19    &10.43   &5.83    &10.59   &5.69    &11.46   &6.06    \\
Fe~I  & 5339.95 & 3.27 & $-$0.72 &13.89   &5.80    &$\cdots$&$\cdots$&$\cdots$&$\cdots$&$\cdots$&$\cdots$\\
Fe~I  & 5379.57 & 3.68 & $-$1.51 &$\cdots$&$\cdots$&$\cdots$&$\cdots$&$\cdots$&$\cdots$&$\cdots$&$\cdots$\\
Fe~I  & 5383.39 & 4.31 &    0.65 &$\cdots$&$\cdots$&$\cdots$&$\cdots$&$\cdots$&$\cdots$&$\cdots$&$\cdots$\\
Fe~I  & 5569.63 & 3.42 & $-$0.49 &12.95   &5.59    &$\cdots$&$\cdots$&$\cdots$&$\cdots$&12.21   &5.88    \\
Fe~I  & 5586.77 & 3.37 & $-$0.12 &$\cdots$&$\cdots$&13.99   &5.79    &$\cdots$&$\cdots$&$\cdots$&$\cdots$\\
Fe~I  & 5859.59 & 4.55 & $-$0.42 &$\cdots$&$\cdots$&$\cdots$&$\cdots$&$\cdots$&$\cdots$&$\cdots$&$\cdots$\\
Fe~I  & 5862.37 & 4.55 & $-$0.13 &$\cdots$&$\cdots$&$\cdots$&$\cdots$&$\cdots$&$\cdots$&$\cdots$&$\cdots$\\
Fe~I  & 5916.26 & 2.45 & $-$2.83 &9.37    &6.09    &$\cdots$&$\cdots$&7.97    &6.36    &$\cdots$&$\cdots$\\
Fe~I  & 5927.79 & 4.65 & $-$1.09 &$\cdots$&$\cdots$&$\cdots$&$\cdots$&$\cdots$&$\cdots$&$\cdots$&$\cdots$\\
Fe~I  & 5934.67 & 3.93 & $-$1.17 &8.92    &6.29    &$\cdots$&$\cdots$&9.10    &6.50    &$\cdots$&$\cdots$\\
Fe~I  & 5956.71 & 0.86 & $-$4.61 &15.11   &6.50    &$\cdots$&$\cdots$&$\cdots$&$\cdots$&$\cdots$&$\cdots$\\
Fe~I  & 5976.79 & 3.94 & $-$1.31 &10.24   &6.63    &$\cdots$&$\cdots$&9.38    &6.62    &$\cdots$&$\cdots$\\
Fe~I  & 6003.02 & 3.88 & $-$1.12 &$\cdots$&$\cdots$&$\cdots$&$\cdots$&$\cdots$&$\cdots$&$\cdots$&$\cdots$\\
Fe~I  & 6008.55 & 3.88 & $-$0.99 &$\cdots$&$\cdots$&$\cdots$&$\cdots$&$\cdots$&$\cdots$&$\cdots$&$\cdots$\\
Fe~I  & 6024.06 & 4.55 & $-$0.12 &9.18    &6.07    &$\cdots$&$\cdots$&$\cdots$&$\cdots$&$\cdots$&$\cdots$\\
Fe~I  & 6027.05 & 4.08 & $-$1.09 &6.94    &6.13    &$\cdots$&$\cdots$&4.65    &6.01    &$\cdots$&$\cdots$\\
Fe~I  & 6056.02 & 4.73 & $-$0.46 &$\cdots$&$\cdots$&$\cdots$&$\cdots$&$\cdots$&$\cdots$&$\cdots$&$\cdots$\\
Fe~I  & 6065.49 & 2.61 & $-$1.53 &$\cdots$&$\cdots$&$\cdots$&$\cdots$&$\cdots$&$\cdots$&$\cdots$&$\cdots$\\
Fe~I  & 6082.72 & 2.22 & $-$3.57 &$\cdots$&$\cdots$&$\cdots$&$\cdots$&$\cdots$&$\cdots$&$\cdots$&$\cdots$\\
Fe~I  & 6151.62 & 2.18 & $-$3.30 &$\cdots$&$\cdots$&$\cdots$&$\cdots$&$\cdots$&$\cdots$&$\cdots$&$\cdots$\\
Fe~I  & 6165.36 & 4.14 & $-$1.47 &$\cdots$&$\cdots$&$\cdots$&$\cdots$&$\cdots$&$\cdots$&$\cdots$&$\cdots$\\
Fe~I  & 6173.33 & 2.22 & $-$2.88 &12.68   &6.25    &10.48   &6.38    &9.48    &6.11    &$\cdots$&$\cdots$\\
Fe~I  & 6180.21 & 2.73 & $-$2.59 &8.11    &6.02    &$\cdots$&$\cdots$&$\cdots$&$\cdots$&$\cdots$&$\cdots$\\
Fe~I  & 6219.29 & 2.20 & $-$2.43 &15.99   &6.25    &12.69   &6.26    &11.84   &5.94    &11.98   &6.19    \\
Fe~I  & 6229.25 & 2.83 & $-$2.81 &$\cdots$&$\cdots$&$\cdots$&$\cdots$&$\cdots$&$\cdots$&$\cdots$&$\cdots$\\
Fe~I  & 6240.67 & 2.22 & $-$3.23 &$\cdots$&$\cdots$&$\cdots$&$\cdots$&9.50    &6.47    &$\cdots$&$\cdots$\\
Fe~I  & 6246.34 & 3.60 & $-$0.73 &$\cdots$&$\cdots$&$\cdots$&$\cdots$&6.85    &5.36    &12.98   &6.40    \\
Fe~I  & 6252.58 & 2.40 & $-$1.69 &$\cdots$&$\cdots$&$\cdots$&$\cdots$&$\cdots$&$\cdots$&12.68   &5.82    \\
Fe~I  & 6254.27 & 2.28 & $-$2.44 &$\cdots$&$\cdots$&10.11   &5.96    &$\cdots$&$\cdots$&$\cdots$&$\cdots$\\
Fe~I  & 6265.15 & 2.18 & $-$2.55 &17.09   &6.46    &$\cdots$&$\cdots$&$\cdots$&$\cdots$&$\cdots$&$\cdots$\\
Fe~I  & 6270.24 & 2.86 & $-$2.46 &8.64    &5.66    &$\cdots$&$\cdots$&$\cdots$&$\cdots$&$\cdots$&$\cdots$\\
Fe~I  & 6301.51 & 3.65 & $-$0.72 &$\cdots$&$\cdots$&$\cdots$&$\cdots$&16.62   &6.74    &$\cdots$&$\cdots$\\
Fe~I  & 6322.69 & 2.59 & $-$2.43 &13.53   &6.40    &$\cdots$&$\cdots$&$\cdots$&$\cdots$&9.17    &6.24    \\
Fe~I  & 6344.15 & 2.43 & $-$2.92 &11.61   &6.41    &$\cdots$&$\cdots$&$\cdots$&$\cdots$&$\cdots$&$\cdots$\\
Fe~I  & 6355.03 & 2.84 & $-$2.35 &11.13   &6.31    &5.47    &5.88    &7.96    &6.17    &$\cdots$&$\cdots$\\
Fe~I  & 6358.71 & 0.86 & $-$4.47 &12.27   &5.91    &$\cdots$&$\cdots$&$\cdots$&$\cdots$&$\cdots$&$\cdots$\\
Fe~I  & 6380.77 & 4.19 & $-$1.38 &$\cdots$&$\cdots$&$\cdots$&$\cdots$&$\cdots$&$\cdots$&$\cdots$&$\cdots$\\
Fe~I  & 6393.62 & 2.43 & $-$1.43 &$\cdots$&$\cdots$&$\cdots$&$\cdots$&$\cdots$&$\cdots$&$\cdots$&$\cdots$\\
Fe~I  & 6411.66 & 3.65 & $-$0.60 &$\cdots$&$\cdots$&10.45   &5.89    &13.80   &6.19    &$\cdots$&$\cdots$\\
Fe~I  & 6419.97 & 4.73 & $-$0.24 &$\cdots$&$\cdots$&8.40    &5.57    &$\cdots$&$\cdots$&$\cdots$&$\cdots$\\
Fe~I  & 6421.37 & 2.28 & $-$2.03 &$\cdots$&$\cdots$&10.88   &5.65    &15.06   &6.08    &$\cdots$&$\cdots$\\
Fe~I  & 6430.86 & 2.18 & $-$2.01 &$\cdots$&$\cdots$&16.13   &6.31    &11.95   &5.48    &15.16   &6.19    \\
Fe~I  & 6475.63 & 2.52 & $-$2.94 &11.29   &6.54    &$\cdots$&$\cdots$&10.61   &6.72    &$\cdots$&$\cdots$\\
Fe~I  & 6518.37 & 2.83 & $-$2.46 &9.88    &6.22    &4.54    &5.83    &3.38    &5.65    &$\cdots$&$\cdots$\\
Fe~I  & 6546.25 & 2.76 & $-$1.54 &$\cdots$&$\cdots$&$\cdots$&$\cdots$&14.64   &6.11    &$\cdots$&$\cdots$\\
Fe~I  & 6569.23 & 4.73 & $-$0.42 &$\cdots$&$\cdots$&$\cdots$&$\cdots$&$\cdots$&$\cdots$&$\cdots$&$\cdots$\\
Fe~I  & 6581.21 & 1.48 & $-$4.68 &4.78    &6.05    &$\cdots$&$\cdots$&$\cdots$&$\cdots$&$\cdots$&$\cdots$\\
Fe~I  & 6593.88 & 2.43 & $-$2.42 &$\cdots$&$\cdots$&$\cdots$&$\cdots$&10.56   &6.03    &$\cdots$&$\cdots$\\
Fe~I  & 6608.04 & 2.28 & $-$4.03 &$\cdots$&$\cdots$&$\cdots$&$\cdots$&$\cdots$&$\cdots$&$\cdots$&$\cdots$\\
Fe~I  & 6677.98 & 2.69 & $-$1.42 &$\cdots$&$\cdots$&$\cdots$&$\cdots$&$\cdots$&$\cdots$&$\cdots$&$\cdots$\\
Fe~I  & 6703.56 & 2.76 & $-$3.16 &$\cdots$&$\cdots$&$\cdots$&$\cdots$&$\cdots$&$\cdots$&$\cdots$&$\cdots$\\
Fe~I  & 6710.34 & 1.49 & $-$4.88 &$\cdots$&$\cdots$&$\cdots$&$\cdots$&$\cdots$&$\cdots$&$\cdots$&$\cdots$\\
Fe~I  & 6750.18 & 2.42 & $-$2.62 &15.06   &6.53    &$\cdots$&$\cdots$&$\cdots$&$\cdots$&$\cdots$&$\cdots$\\
Fe~II & 4993.36 & 2.81 & $-$3.62 &$\cdots$&$\cdots$&$\cdots$&$\cdots$&$\cdots$&$\cdots$&$\cdots$&$\cdots$\\
Fe~II & 5234.65 & 3.22 & $-$2.18 &$\cdots$&$\cdots$&$\cdots$&$\cdots$&$\cdots$&$\cdots$&11.51   &6.33    \\
Fe~II & 5425.25 & 3.20 & $-$3.22 &7.30    &6.72    &$\cdots$&$\cdots$&$\cdots$&$\cdots$&$\cdots$&$\cdots$\\
Fe~II & 6247.60 & 3.89 & $-$2.30 &$\cdots$&$\cdots$&4.98    &6.17    &$\cdots$&$\cdots$&$\cdots$&$\cdots$\\
Fe~II & 6432.68 & 2.89 & $-$3.57 &8.24    &6.78    &$\cdots$&$\cdots$&5.23    &6.25    &5.28    &6.30    \\
Fe~II & 6456.44 & 3.90 & $-$2.05 &$\cdots$&$\cdots$&$\cdots$&$\cdots$&5.66    &5.99    &$\cdots$&$\cdots$\\
Fe~II & 6516.10 & 2.89 & $-$3.31 &7.52    &6.35    &5.84    &6.12    &$\cdots$&$\cdots$&$\cdots$&$\cdots$\\
Co~I  & 5301.04 & 1.71 & $-$2.00 &$\cdots$&$\cdots$&$\cdots$&$\cdots$&$\cdots$&$\cdots$&$\cdots$&$\cdots$\\
Co~I  & 5647.23 & 2.28 & $-$1.56 &$\cdots$&$\cdots$&$\cdots$&$\cdots$&$\cdots$&$\cdots$&$\cdots$&$\cdots$\\
Co~I  & 5991.86 & 2.08 & $-$1.85 &$\cdots$&$\cdots$&$\cdots$&$\cdots$&$\cdots$&$\cdots$&$\cdots$&$\cdots$\\
Co~I  & 6771.05 & 1.88 & $-$1.97 &$\cdots$&$\cdots$&$\cdots$&$\cdots$&$\cdots$&$\cdots$&$\cdots$&$\cdots$\\
Ni~I  & 4904.42 & 3.54 & $-$0.17 &$\cdots$&$\cdots$&$\cdots$&$\cdots$&$\cdots$&$\cdots$&$\cdots$&$\cdots$\\
Ni~I  & 5035.39 & 3.63 &    0.29 &$\cdots$&$\cdots$&$\cdots$&$\cdots$&$\cdots$&$\cdots$&$\cdots$&$\cdots$\\
Ni~I  & 5080.53 & 3.65 &    0.13 &11.91   &4.98    &$\cdots$&$\cdots$&$\cdots$&$\cdots$&$\cdots$&$\cdots$\\
Ni~I  & 5084.11 & 3.68 &    0.03 &$\cdots$&$\cdots$&$\cdots$&$\cdots$&$\cdots$&$\cdots$&6.26    &4.54    \\
Ni~I  & 5146.48 & 3.71 & $-$0.06 &$\cdots$&$\cdots$&9.28    &5.12    &$\cdots$&$\cdots$&$\cdots$&$\cdots$\\
Ni~I  & 5578.73 & 1.68 & $-$2.64 &$\cdots$&$\cdots$&$\cdots$&$\cdots$&$\cdots$&$\cdots$&$\cdots$&$\cdots$\\
Ni~I  & 5587.85 & 1.94 & $-$2.14 &$\cdots$&$\cdots$&$\cdots$&$\cdots$&$\cdots$&$\cdots$&$\cdots$&$\cdots$\\
Ni~I  & 5592.29 & 1.95 & $-$2.59 &$\cdots$&$\cdots$&$\cdots$&$\cdots$&$\cdots$&$\cdots$&$\cdots$&$\cdots$\\
Ni~I  & 5846.99 & 1.68 & $-$3.21 &$\cdots$&$\cdots$&$\cdots$&$\cdots$&$\cdots$&$\cdots$&$\cdots$&$\cdots$\\
Ni~I  & 6108.12 & 1.68 & $-$2.45 &$\cdots$&$\cdots$&9.13    &4.84    &8.84    &4.70    &$\cdots$&$\cdots$\\
Ni~I  & 6128.98 & 1.68 & $-$3.33 &$\cdots$&$\cdots$&$\cdots$&$\cdots$&$\cdots$&$\cdots$&$\cdots$&$\cdots$\\
Ni~I  & 6176.80 & 4.09 & $-$0.43 &$\cdots$&$\cdots$&$\cdots$&$\cdots$&$\cdots$&$\cdots$&$\cdots$&$\cdots$\\
Ni~I  & 6177.25 & 1.83 & $-$3.60 &$\cdots$&$\cdots$&$\cdots$&$\cdots$&$\cdots$&$\cdots$&$\cdots$&$\cdots$\\
Ni~I  & 6327.59 & 1.68 & $-$3.15 &9.12    &5.07    &$\cdots$&$\cdots$&$\cdots$&$\cdots$&$\cdots$&$\cdots$\\
Ni~I  & 6482.79 & 1.94 & $-$2.63 &8.59    &4.88    &$\cdots$&$\cdots$&$\cdots$&$\cdots$&$\cdots$&$\cdots$\\
Ni~I  & 6532.90 & 1.94 & $-$3.39 &$\cdots$&$\cdots$&$\cdots$&$\cdots$&$\cdots$&$\cdots$&$\cdots$&$\cdots$\\
Ni~I  & 6586.32 & 1.95 & $-$2.81 &$\cdots$&$\cdots$&4.47    &4.83    &$\cdots$&$\cdots$&$\cdots$&$\cdots$\\
Ni~I  & 6643.64 & 1.68 & $-$2.30 &$\cdots$&$\cdots$&$\cdots$&$\cdots$&13.24   &5.07    &9.97    &4.84    \\
Ni~I  & 6767.78 & 1.83 & $-$2.17 &11.57   &4.63    &$\cdots$&$\cdots$&$\cdots$&$\cdots$&7.38    &4.53    \\
Ni~I  & 6772.30 & 3.66 & $-$0.98 &4.32    &4.90    &$\cdots$&$\cdots$&6.86    &5.41    &$\cdots$&$\cdots$\\
Cu~I  & 5105.55 & 1.39 & $-$1.51 &$\cdots$&$\cdots$&syn     &2.39    &$\cdots$&$\cdots$&$\cdots$&$\cdots$\\
Y~II  & 5087.44 & 1.08 & $-$0.16 &9.52    &0.67    &$\cdots$&$\cdots$&$\cdots$&$\cdots$&$\cdots$&$\cdots$\\
Y~II  & 5200.44 & 0.99 & $-$0.57 &$\cdots$&$\cdots$&$\cdots$&$\cdots$&$\cdots$&$\cdots$&7.5     &0.94    \\
Y~II  & 5509.94 & 0.99 & $-$1.01 &4.73    &0.72    &7.65    &1.34    &$\cdots$&$\cdots$&$\cdots$&$\cdots$\\
Zr~I  & 6127.47 & 0.15 & $-$1.06 &$\cdots$&$\cdots$&$\cdots$&$\cdots$&$\cdots$&$\cdots$&$\cdots$&$\cdots$\\
Zr~I  & 6134.55 & 0.00 & $-$1.28 &$\cdots$&$\cdots$&$\cdots$&$\cdots$&$\cdots$&$\cdots$&$\cdots$&$\cdots$\\
Zr~II & 5112.32 & 1.66 & $-$0.59 &$\cdots$&$\cdots$&$\cdots$&$\cdots$&$\cdots$&$\cdots$&$\cdots$&$\cdots$\\
Ba~II & 5853.70 & 0.60 & $-$1.01 &13.86   &0.90    &14.79   &1.52    &$\cdots$&$\cdots$&$\cdots$&$\cdots$\\
Ba~II & 6141.72 & 0.70 & $-$0.08 &$\cdots$&$\cdots$&syn     &0.93    &17.92   &0.79    &17.45   &1.00    \\
Ba~II & 6496.92 & 0.60 & $-$0.38 &syn     &0.83    &16.54   &1.03    &16.42   &0.69    &12.73   &0.39    \\
La~II & 6320.41 & 0.17 & $-$1.56 &$\cdots$&$\cdots$&$\cdots$&$\cdots$&$\cdots$&$\cdots$&$\cdots$&$\cdots$\\
La~II & 6390.50 & 4.15 & $-$1.40 &6.97    &0.75    &$\cdots$&$\cdots$&$\cdots$&$\cdots$&$\cdots$&$\cdots$\\
Ce~II & 5274.26 & 1.04 &    0.15 &4.82    &0.64    &$\cdots$&$\cdots$&$\cdots$&$\cdots$&$\cdots$&$\cdots$\\
Nd~II & 5249.58 & 0.98 &    0.22 &$\cdots$&$\cdots$&$\cdots$&$\cdots$&$\cdots$&$\cdots$&$\cdots$&$\cdots$\\
Nd~II & 5319.82 & 0.55 & $-$0.19 &7.58    &0.46    &7.49    &0.72    &$\cdots$&$\cdots$&8.64    &0.95    \\
Eu~II & 6645.12 & 1.38 &    0.12 &$\cdots$&$\cdots$&$\cdots$&$\cdots$&$\cdots$&$\cdots$&$\cdots$&$\cdots$\\
Dy~II & 5169.69 & 0.10 & $-$1.66&$\cdots$&$\cdots$&$\cdots$&$\cdots$&$\cdots$&$\cdots$&$\cdots$&$\cdots$\\                       
\end{longtable}                                                                                                                   
}                                                                                                                                 
\end{appendix}                                                                                                                    
\end{document}